\documentclass[12pt]{article}

\usepackage{latexsym}  
\usepackage{graphicx}  

\begin{document}

\baselineskip=0.5truecm
\title{
A Tutorial on Advanced Dynamic Monte Carlo Methods
for Systems with Discrete State Spaces
}
\author{
\bf{M.~A.\ Novotny}\\
\\
School of Computational Science and Information Technology, \\
Florida State University, Tallahassee, FL 32306-4120, U.S.A.\\
\\
\\
Permanent Address: Dept.\ of Physics and Astronomy\\
Mississippi State University\\
Mississippi State, MS ~ 39762, USA \\
}
\maketitle
\begin{abstract}
Advanced algorithms are necessary to obtain faster-than-real-time 
dynamic simulations in a number of different physical problems that 
are characterized by widely disparate time scales.  
Recent advanced dynamic Monte Carlo algorithms that preserve the 
dynamics of the model are described.  
These include the $n$-fold way 
algorithm, the Monte Carlo with Absorbing Markov Chains (MCAMC) algorithm, 
and the Projective Dynamics (PD) algorithm.  
To demonstrate the use of these algorithms, they are applied to some 
simplified models of dynamic physical systems.  
The models studied include a model for ion motion through a pore such as 
a biological ion channel and 
the metastable decay of the ferromagnetic Ising model.  
Non-trivial parallelization issues for these dynamic algorithms,
which are in the class of parallel discrete event simulations,
are discussed.  
Efforts are made to keep the article 
at an elementary level by concentrating on a simple model 
in each case that illustrates the 
use of the advanced dynamic Monte Carlo algorithm.  

\end{abstract}

\clearpage


\section*{1.0~Introduction}
\label{Sec-Intro}

Classical Monte Carlo methods have been used to study a wide variety of 
different systems since they were first proposed by 
Metropolis, Rosenbluth, Rosenbluth, Teller, and Teller \cite{METR53}.  
The Monte Carlo method can be viewed as a method to perform 
multi-dimensional integrals \cite{PRES86}.  For example, to 
obtain static critical exponents \cite{STAN71} 
for model systems such as Ising, Potts, and 
classical Heisenberg  
models,  
Monte Carlo simulations are used to calculate multidimensional sums or 
integrals \cite{BIND86,BIND97}.    
A significant number of advanced Monte Carlo 
algorithms have been developed 
for static critical phenomena.  These include 
cluster algorithms 
[6--8], 
multicanonical algorithms 
[9--15], 
and histogram reweighting methods 
[16--18] 

In its simplest form, the Monte Carlo algorithm is 
like a movie with each frame a three-step process.
First, generate a (pseudo-)random number.   
In this work we will only use random 
numbers uniformly distributed between 0 and 1.  
Second, choose a trial move from the current state to a new state.  
Third, accept or reject this trial move depending on the current random 
number and an acceptance probability for the trial move.  
The sums or integrals to be calculated are then updated using 
the new frame, and the 
procedure is iterated repeatedly.  

Calculations of static behaviors, such as multi-dimensional 
integrals or static critical 
exponents, allow a wide variety of trial moves.  This includes, for 
example, cluster and multicanonical algorithms.  This is so 
because in 
static calculations only the final sum (or integral) is of interest.  
The particular trial moves have no physical meaning.  Such freedom of 
trial moves no longer exists if the 
Monte Carlo algorithm is applied to a dynamic problem where the 
physical meaning of the trial moves is an essential part of the 
model.  Fortunately, some advanced Monte Carlo algorithms exist to make 
many of these calculations efficient without changing the meaning of 
the trial moves.  It is these advanced algorithms that will be 
described and applied in this article.  


\section*{2.0~Standard Dynamic Monte Carlo Methods}
\label{Sec-MC}

It is also possible to ascribe physical meaning to certain trial moves.  
Consider a one-dimensional random walker on a lattice.  At each 
tick of a clock the walker has a certain probability to move to the 
left, the right, or to stay where it is.  
In this case, the actual Monte Carlo trial (move left, move right, or stay) 
has a physical meaning.  
Now imagine that the probabilities of moving to the left or 
to the right are small.  
Then, most of the time the walker stays where it is, 
and it will require a large amount of 
computer time before the walker moves a substantial distance.  The advanced 
dynamic algorithms for this problem, described in Sec.~4, 
keep this physical dynamic intact, 
but decrease the amount of computer time required for 
the walker to reach a specified location.  

Consider the ferromagnetic Ising model on a regular lattice with 
periodic boundary conditions.  Each site of the lattice has a spin 
$\sigma_i$$=$$\pm$$1$.  
From the Hamiltonian, the energy of a particular spin configuration 
is given by 
\begin{equation}
\label{EHIsing} 
{\cal E} = 
-J \sum_{\langle i,j\rangle} \sigma_i \sigma_j - H \sum_i \sigma_i
, \; 
\end{equation}
where the coupling constant $J$$>$$0$ and the 
external magnetic field is $H$.  
The first sum of Eq.~(\ref{EHIsing}) is over all nearest-neighbor 
spin pairs and the second sum is over all $N$ Ising spins.  
One can use the simple Monte Carlo procedure described in this section 
or advanced methods such as a cluster method,   
to obtain the static critical behavior, such as 
the internal energy, specific heat, susceptibility, or order parameter at 
a particular temperature $T$ and field $H$.  
A particularly simple trial move involves 
a single spin flip, i.e.\ letting 
$\sigma_i$$\rightarrow$$-\sigma_i$.  
However, one can also consider these 
Monte Carlo trial moves to be due to 
the interaction of the Ising spins with a heat bath.  Then the dynamics 
of the simulation is of physical relevance, and methods such as the 
cluster method {\it cannot\/} be used since they change the dynamics.  

The dynamics for the kinetic Ising model can be derived using 
a Master equation approach \cite{GLAU63,BIND73,BIND74}.  
In 1977 Martin \cite{MART77} considered a lattice of quantum 
spin ${1\over 2}$ particles, each coupled to their own heat bath, 
and showed that under 
certain assumptions the same dynamic as for the kinetic Ising model 
was obtained. 
This dynamic is: 
\begin{itemize}
\item Randomly choose an Ising spin.  If there are $N$ Ising spins, the 
probability of choosing each spin is $1/N$.  
\item Calculate a random number ${\bar r}$.
\item Calculate (or look up in a stored table) 
the energy $E_{\rm old}$ of the current Ising configuration 
and $E_{\rm new}$ of the configuration with the chosen Ising spin flipped.  
\item Change to the new configuration, i.e.\ flip the chosen Ising spin, 
if 
\begin{equation}
\label{EIGlauber}
{\bar r}\le
\exp(-E_{\rm new}/k_{\rm B}T) / 
\left[\exp(-E_{\rm old}/k_{\rm B}T)+\exp(-E_{\rm new}/k_{\rm B}T)\right]
\; .  
\end{equation}
\item Increment the time by one Monte Carlo step (mcs), 
i.e.\ by one spin-flip attempt.  
\end{itemize} 
This Monte Carlo dynamic has physical meaning.  
Here $T$ is the temperature and 
$k_{\rm B}$ is Boltzmann's constant.  
Furthermore, 
the time scale for a single spin-flip trial is set by the heat bath.  
This means that for Ising spins interacting with phonons, one 
spin-flip attempt per spin (one Monte Carlo step per spin, 
defined to be one MCSS) should correspond to a 
time on the order of an inverse phonon frequency, 
roughly $10^{-13}$ seconds.  Thus, to simulate this simple model 
for a time of one second would take $10^{13}$ Monte Carlo steps per 
spin.  
If you could get your computer to perform one spin-flip attempt 
for each CPU clock tick 
(roughly $10^{-9}$ sec), 
then a one-second simulation would take about three hours of CPU time 
per spin!  
Your simulation would take $10^4$$N$ times longer than physical time.  
To simulate metastable decay for the Ising model, as discussed below 
in Sec.~4.1, 
one is interested in 
time scales of years to decades.  
To simulate your Ising model for a $N$$=$$100$ spin system 
would take about 12 days of CPU time to simulate for one second, 
about 2 years of CPU time to simulate for one minute, and 
about one century of CPU time to simulate for one hour.  
These CPU times are much 
longer than I can wait for computer results, 
and we are interested in simulating much larger lattices and 
much longer times.  
The sections below 
describe advanced dynamic algorithms that allow these simulations to be 
performed in a reasonable amount of CPU time {\it without\/} changing 
the underlying dynamic.  The goal is to obtain faster-than-real-time 
dynamic simulations.  

The dynamic algorithm described above, 
using Eq.~(\ref{EIGlauber}) for the Ising model, is called the 
Glauber dynamic.  
It was introduced by Glauber \cite{GLAU63} to 
study the dynamics of a one-dimensional Ising chain.  
For higher-spin models, and sometimes also for the 
Ising model, this type of dynamic is called a 
heat-bath dynamic.  
Another popular choice \cite{BIND86,BIND97} 
for the flip probability for Ising spins is 
the Metropolis dynamic \cite{METR53}, 
which always flips the spin if the energy of the 
new configuration is lower than that of the old.  
Thus the spin flips if 
\begin{equation}
\label{EIMetrop}
{\bar r} \le 
{\rm min}\left\{ 1, 
\exp\left[(E_{\rm old}-E_{\rm new})/k_{\rm B} T \right]\right\}
\; .
\end{equation}
The dynamic 
that should be used in a particular dynamic simulation 
is the one that has been derived from the underlying 
physical system.  Both of these dynamics satisfy detailed 
balance, and hence will obtain the correct statics for 
the Ising model.  

Note that the dynamic involves a {\it random\/} choice of 
the spin on which a spin flip will be attempted.  Performing 
spin-flip attempts sequentially on the lattice will lead to a different 
dynamic.  It has been shown in nucleation-and-growth studies of 
the lifetime of the metastable state in an Ising model that the 
use of sequential instead of random updates can lead to different physical 
results.  In particular, the pre-factor for the nucleation rate is 
different \cite{RIKV94A,RIKV94B}.  


\section*{3.0~A Brief Review of Absorbing Markov Chains}
\label{Sec-AMC}

Many of the advanced algorithms described in the following sections will 
use absorbing Markov chains.  A brief introduction to some of the 
important features of absorbing Markov chains is presented here.  

Consider an absorbing Markov chain with $s$ transient states and $r$ 
absorbing states.  The system starts in one of the $s$ transient states, 
and remains in the transient subspace of the $s$ transient states until 
it is absorbed into one of the $r$ absorbing states.  
We will use the mathematician's notation of the current state vector being 
operated on by a matrix from the right.  
(As opposed to the normal physicist's 
convention.)  This will allow the reader to make easy 
contact with the literature on absorbing 
Markov chains, for example \cite{IOSI80,STEW94}.  

The transition matrix defining 
an absorbing Markov chain has the form
\begin{equation}
\label{EAMC0} 
{\bf M}_{(r+s)\times(r+s)} = 
\pmatrix{
{\bf I}_{r\times r} & {\bf 0}_{r\times s} \cr
{\bf R}_{s\times r} & {\bf T}_{s\times s} \cr
}
\end{equation}
where the sizes of the matrices have been explicitly shown.  The matrix 
${\bf I}$ is an identity matrix, the matrix ${\bf 0}$ is a matrix with 
all zero elements, the matrix ${\bf T}$ is called the transient matrix, 
and the matrix ${\bf R}$ is called the recurrent matrix.  
${\bf M}$ is a Markov matrix, i.e., each of its row sums equal one, 
since at each clock tick the system must jump to a new state or 
remain in the current state.  

The matrix that governs the motion of the system after $m$ time steps (or 
$m$ clock ticks) is given by matrix multiplication to be 
\begin{equation}
\label{EAMC1}
{\bf M}^m_{(r+s)\times(r+s)} =
\pmatrix{
{\bf I}_{r\times r} & {\bf 0}_{r\times s} \cr
({\bf I}+{\bf T}+\cdots+{\bf T}^{m-1})_{s\times s}{\bf R}_{s\times r} 
& {\bf T}^m_{s\times s} \cr
}
\; .
\end{equation}
The system must initially be in the transient subspace, so the $s$-state 
initial vector is ${\vec v}_{\rm I}^{\rm T}$ and the total initial vector for 
the matrix ${\bf M}$ is given by 
$\pmatrix{ 
{\vec 0}^{\rm T} & {\vec v}_{\rm I}^{\rm T}
}$
where the vector of zeros has $r$ elements.  
(The superscript ${\rm T}$ denotes the transponse.)  
Applying ${\bf M}^m$ to this 
initial vector gives the $(r$$+$$s)$-dimensional vector 
\begin{equation}
\label{EAMC2}
\pmatrix{ {\vec 0}^{\rm T} & {\vec v}_{\rm I}^{\rm T} } 
{\bf M}^m=
\pmatrix{
{\vec v}_{\rm I}^{\rm T} 
\left ({\bf I}+{\bf T}+\cdots+{\bf T}^{m-1}\right ) {\bf R} 
&
{\vec v}_{\rm I}^{\rm T} {\bf T}^m 
}
\; .
\end{equation}
Introduce the vector ${\vec e}$ with all elements equal to unity.  Then the 
probability of still being in the space of the 
$s$ transient states after $m$ time steps 
is given by 
\begin{equation}
\label{EAMC3}
p_{\rm still\ in\ transient\ subspace} = 
{\vec v}_{\rm I}^{\rm T} {\bf T}^m {\vec e}
\; .
\end{equation}
The probability of the state having exited to each of the 
$r$ absorbing states after $m$ 
time steps is given by the elements of the vector 
\begin{equation}
\label{EAMC4}
{\vec p}^{\rm T}_{{\rm absorption\ after\ }m{\rm \ time\ steps}} = 
{\vec v}_{\rm I}^{\rm T} 
\left({\bf I}+{\bf T}+\cdots+{\bf T}^{m-1}\right) {\bf R}
\; .
\end{equation}
The probability of having exited to each of the $r$ absorbing states 
after $m$ 
time steps is the sum of the probabilities of 
having exited in the first time step (which is 
the term ${\bf I}{\bf R}$) plus the probability of having 
exited after the second 
time step (the term ${\bf T}{\bf R}$) plus the probability of having exited 
in time steps 3, 4, $\cdots$, $m$.  Thus the probability that the system 
exits to each of the $r$ absorbing states, {\it given\/} that 
it exits at time 
step $m$, is given by the elements of the vector 
\begin{equation}
\label{EAMC5} 
{\vec p}^{\rm T}_{{\rm absorption\ given\ exit\ at\ time\ }m} = 
{\vec q} = 
{
{ {\vec v}_{\rm I}^{\rm T} {\bf T}^{m-1} {\bf R} } \over
{ {\vec v}_{\rm I}^{\rm T} {\bf T}^{m-1} {\bf R} {\vec e} } }
\; .
\end{equation}
Equations (\ref{EAMC3}) and (\ref{EAMC5}) 
are the equations needed to perform 
Monte Carlo with Absorbing Markov Chain calculations.  

In the absorbing Markov chain literature the $s$$\times$$s$ matrix 
\begin{equation}
\label{EAMC6} 
{\bf N} = \left({\bf I}-{\bf T}\right)^{-1} 
= {\bf I} + {\bf T} + {\bf T}^2 + \cdots
\; ,
\end{equation}
called the fundamental matrix, is introduced.  
Then the probability that the system ends up in a particular 
absorbing state is given by
the elements of the vector 
\begin{equation}
\label{EAMC7}
{\vec p}^{\rm T}_{\rm absorbed\ into\ particular\ state} = 
{\vec v}_{\rm I}^{\rm T} {\bf N} {\bf R}
\; .
\end{equation}
The average lifetime to exit from the transient subspace is 
given by
\begin{equation}
\label{EAMC8} 
\langle\tau\rangle
=
{\vec v}_{\rm I}^{\rm T} {\bf N} {\vec e} 
\; .  
\end{equation}
The units for $\langle\tau\rangle$ are the time units of the clock, 
with one attempted move every tick of the clock.  
This can be derived using the fact 
that the probability to be absorbed at time step 
$m$ is 
\begin{equation}
\label{EAMC9} 
p_{{\rm absorbed\ at\ time\ } m}
=
{\vec v}_{\rm I}^{\rm T} \left ( 
{\bf T}^{m-1} - {\bf T}^m \right ) {\vec e} 
\; ,
\end{equation}
so that 
\begin{eqnarray}
\label{EAMCtau} 
\langle\tau\rangle  & = &
\sum_{m=1}^\infty 
m 
{\vec v}_{\rm I}^{\rm T} \left ( 
{\bf T}^{m-1} - {\bf T}^m \right ) {\vec e} \nonumber \\
& = &
{\vec v}_{\rm I}^{\rm T} \left ( 
{\bf I} + {\bf T}^1 + {\bf T}^2 + \cdots \right ) {\vec e} \nonumber \\
& = & 
{\vec v}_{\rm I}^{\rm T} \left ( 
{\bf I} -{\bf T} \right )^{-1} {\vec e} 
\> = \>
{\vec v}_{\rm I}^{\rm T} {\bf N} {\vec e} 
\; .
\end{eqnarray}
The fundamental matrix can also be used to 
calculate the moments of the lifetime for 
the system, that is the moments of the exit time probabilities.  
These are given by \cite{IOSI80} 
\begin{equation}
\label{EAMCtau2} 
\left\langle\tau^2\right\rangle = 
{\vec v}_{\rm I}^{\rm T} \left( 
2{\bf N}^2 - {\bf N} 
\right ) {\vec e} 
\end{equation}
and 
\begin{equation}
\label{EAMCtau3} 
\left\langle\tau^3\right\rangle = 
{\vec v}_{\rm I}^{\rm T} \left( 
6{\bf N}^3 -6{\bf N}^2 + {\bf N} 
\right ) {\vec e} 
\end{equation}
and 
\begin{equation}
\label{EAMCtau4} 
\left\langle\tau^4\right\rangle = 
{\vec v}_{\rm I}^{\rm T} \left( 
24{\bf N}^4-36{\bf N}^3 +14{\bf N}^2 - {\bf N} 
\right ) {\vec e} 
\; .
\end{equation}


\section*{4.0~Particle Motion Through a Pore}
\label{Sec-Part0}

In order to illustrate the advanced dynamic 
Monte Carlo algorithms, we introduce an 
extremely simple one-dimensional walk.  This one-dimensional walk 
can be viewed as a first step in simulating motion of ions through 
biological ionic channels 
[27--32]. 
These ion channels control the motion of 
ions through cell membranes, and they are consequently of importance 
in understanding cell function and drug delivery to cells.  
The model is kept simple to allow us to introduce the 
dynamic Monte Carlo algorithms 
in a transparent fashion.  

The model is a one-dimensional walk on a lattice.  
Our particle (the ion) starts at 
lattice site zero (on the left) and performs a random walk until it is 
absorbed at the right-most lattice site.  
Each lattice site $i$ 
has an energy $E_i$.  We will study a 16-site lattice 
(16 transient states and one absorbing state) 
with energies shown in 
Fig.~\ref{figIonE}.  
The energies for lattice sites $0$ through $16$ are 
(in dimensionless units) 
$-{1\over 2}$, 
$1$,          $1$,          $0$, ${1\over 2}$, 
$0$,          $0$, ${1\over 2}$,          $0$, 
$0$, ${1\over 2}$, ${1\over 2}$, ${3\over 4}$, 
$0$, ${1\over 2}$,          $0$,         $-1$.  
The dynamic we introduce is that at each time step 
the walker randomly picks whether it will try to perform a move to the 
left or to the right.  The probability of actually performing this move 
is given by
\begin{equation}
\label{p_ion}
p_{\rm move} = 
{{\exp(-E_{i\pm1}/k_{\rm B}T)}\over
{\exp(-E_i/k_{\rm B}T)+\exp(-E_{i\pm1}/k_{\rm B}T)}}
\; ,
\end{equation}
where the $+$ ($-$) sign is for the attempted move to the right (left).  
When the walker is in lattice site~0, the move to the right is always 
attempted.  
We are interested in the average lifetime $\langle\tau\rangle$ 
for a walker 
starting at lattice site~0 
before it is absorbed at lattice site~16.  

\subsection*{4.1~Ion Channel Standard Monte Carlo}
\label{Sec-Part4mc}

The standard Monte Carlo dynamic for this problem is extremely simple to 
program.  Each algorithmic step uses two uniformly distributed random 
numbers, ${\tilde r}$ and ${\bar r}$.  
The algorithm is:
\begin{itemize}
\item Place the walker at lattice site~0.  
\item Set the time $t$ to zero.  
\item Perform the following steps until the walker is absorbed when 
it reaches lattice site~16.
\begin{itemize}
\item[$\circ$] Set the time $t$ to $t$$=$$t$$+$$1$.
\item[$\circ$] Calculate two random numbers, ${\tilde r}$ and ${\bar r}$.
\item[$\circ$] If the walker is not at lattice site~0 and ${\tilde r}$ is 
less than $1\over 2$, then attempt a move to the left.  Otherwise attempt a 
move to the right.  
\item[$\circ$] Calculate (or look up in a stored table) the probability of 
moving, $p_{\rm move}$ from Eq.~(\ref{p_ion}).
\item[$\circ$] Move the walker if 
${\bar r}$$\le$$p_{\rm move}$.  
\end{itemize}
\end{itemize}
This algorithm is executed $K$ times, and the statistics for the 
mean lifetime are accumulated.  
The average lifetime, calculated for $K$$=$$1000$ escapes, is shown in 
Fig.~\ref{figIonTau}.  The time unit is Monte Carlo steps (mcs), 
so the lifetime 
corresponds to the average number of clock ticks before the 
walker reaches lattice site~16.  

\subsection*{4.2~Ion Channel 
$n$-fold Way Monte Carlo (Event Driven Simulation)}
\label{Sec-Part4nfold}

As can be seen from Fig.~\ref{figIonTau}, the average lifetime at 
low temperatures 
can be very long.  
The time unit is Monte Carlo steps, mcs, for our single walker.  
At $T$$=$$0.05$ (in dimensionless units with $k_{\rm B}$$=$$1$) 
the value for $\tau$ is about 
$3$$\times$$10^{13}$~mcs.  If each mcs took a nanosecond of computer time (a 
very fast implementation on an extremely fast computer), the computational 
time required for 1000 escapes would require 
$3$$\times$$10^7$ seconds of CPU time.  This corresponds to about 
one year of computer time!  (For my implementation of the standard 
algorithm on a Cray SV1 supercomputer 
this calculation would have taken about 32 years!)  
I did not spend this amount of computer time on the calculation.  
Rather, I utilized an advanced algorithm to perform my dynamic simulation 
at this temperature in 35 minutes on a Cray SV1 computer.   

The first advanced algorithm I used corresponds to the $n$-fold way 
algorithm introduced by Bortz, Kalos, and Lebowitz in 1975 \cite{BORT75}, 
as implemented in discrete time \cite{NOVO95}.  
This and related similar schemes have been put forward 
and utilized in the simulations of many problems.  
Just a few of these include 
crystal growth \cite{GILM76,MAKS88}, 
renormalization group methods for obtaining critical exponents \cite{MA76}, 
growth of nanostructures \cite{JENS99A,JENS99B}, 
dynamics of simple models of glasses \cite{NEWM99,SOLL99}, 
lattice models for protein folding \cite{MELI99}, 
thermal decay of recorded information \cite{CHAR97}, 
Potts models for domain growth \cite{SAHN83}, 
catalytic surface reactions \cite{KANG95,LUKK98}, 
aging on long time scales \cite{KRAU95}, 
ground state calculations of spin glass models \cite{GRES86},
crossover phenomena in the anisotropic Ising model \cite{GRAI81}, 
domain boundaries in anisotropic Ising models \cite{OZAW90}, 
multidimensional minimization of functions via simulated annealing 
\cite{GREE86}, 
dynamic recrystallization \cite{PECZ92}, 
and 
collective surface diffusion \cite{BULN98}.  

These algorithms 
are also called event-driven algorithms in the 
discrete-event simulation community \cite{LAWK91}.  
A discrete-event simulation is one in which the state vector of the system 
changes discontinuously at particular (possibly random) time intervals.  
In this case, the time intervals correspond to the times at which the 
walker jumps from one state to another ({\it not\/} the times at which 
a step is attempted but then rejected).  
The idea behind event-driven simulations is to change the state of the 
system in one algorithmic step, and to increment 
the time by the time required 
to make that transition.  This is where the absorbing Markov chain methodology 
is useful.  

Let the transient state for an absorbing Markov chain be the current 
state, and the two absorbing states be the walk to the left or to the 
right.  Then the Markov matrix is 
\begin{equation}
{\bf M}_{\rm ion\ 1} = 
\pmatrix{
1 & 0 & 0 \cr
0 & 1 & 0 \cr
p_{\rm left} & p_{\rm right} & 1-p_{\rm left}-p_{\rm right} \cr
}
\end{equation}
where $p_{\rm right}$ is the probability of jumping to the right and 
$p_{\rm left}$ is the probability of jumping to the left.  
These probabilities are given by $p_{\rm move}/2$ 
[from Eq.~(\ref{p_ion})] 
where the factor of 
$1/2$ is needed 
because a move attempt to the left or right is chosen at random.  
If the walker is at lattice site~0, then $p_{\rm left}$$=$$0$ and 
$p_{\rm right}$$=$$p_{\rm move}$.   
Since the walker is in the current state, the initial state vector is 
${\vec v}^{\rm T}_{\rm I} = \pmatrix{0 & 0 & 1\cr}$.   

Each algorithmic step again uses two uniformly distributed random numbers, 
${\tilde r}$ and ${\bar r}$.  
The algorithm to repeat for $K$ escapes is:
\begin{itemize}
\item Place the walker at lattice site~0.
\item Set the time clock to $t$$=$$0$.
\item Perform the following steps until the walker is absorbed at lattice 
site~16.  
\begin{itemize}
\item[$\circ$] Calculate two random numbers, ${\tilde r}$ and ${\bar r}$.  
\item[$\circ$] Calculate (or look up from tables) the two 
probabilities $p_{\rm right}$ and $p_{\rm left}$.  
\item[$\circ$] Calculate the time $m$ to exit from the current state 
$m$$=$$\left\lfloor 
{{\ln({\bar r})}\over{\ln(1-p_{\rm left}-p_{\rm right})}} \right\rfloor + 1$, 
where $\left\lfloor z \right\rfloor$ is the integer part of $z$.  
\item[$\circ$] Move the walker to the right if 
${\tilde r}$$\le$${{p_{\rm right}}\over{p_{\rm right}+p_{\rm left}}}$.  
Otherwise move the walker to the left.  
\item[$\circ$] Increment the time to $t$$+$$m$.  
\end{itemize}
\end{itemize}

In a primitive way, if the probability to move is very small, say 0.001 per
time step, then instead of increasing the time by one unit for each of the 
mostly unsuccessful attempts to move, we always move the particle and 
increase the time by on average 
1000 units, thus saving computing effort by a factor 1000. 

A number of remarks are required.  
The number of time steps $m$ to exit the current state is given by 
Eq.~(\ref{EAMC3}).  For this simple case where there is only 
one transient state ($s$$=$$1$), the matrix 
${\bf T}$$=$$T_{11}$$=$$1$$-$$p_{\rm left}$$-$$p_{\rm right}$ 
is a scalar, 
and 
Eq.~(\ref{EAMC3}) gives the time of exit from the 
transient state to be 
$T_{11}^m$$<$${\bar r}$$\le$$T_{11}^{m-1}$.  
This reduces to the equation 
for $m$ in terms of the natural logarithms.  
Also, since ${\bf T}$ is a 
scalar, there is no $m$ dependence on the probabilities to 
exit to the left or right state since from Eq.~(\ref{EAMC5}) 
the factor of $T_{11}^{m-1}$ cancels due to the normalization.  

Using the event-driven simulation, the time to measure 
$K$$=$$1000$ escapes of 
the algorithm at $T$$=$$0.05$ would be $10^3$ minutes, or about 
17 hours of Cray SV1 time.  
For this temperature the event-driven simulation is about one 
million times faster than the standard Monte Carlo algorithm! 

\subsection*{4.3~Ion Channel $s$$=$$2$ MCAMC}
\label{Sec-Part4mcamc}

\begin{figure}[t]
\includegraphics[width=.60\textwidth]{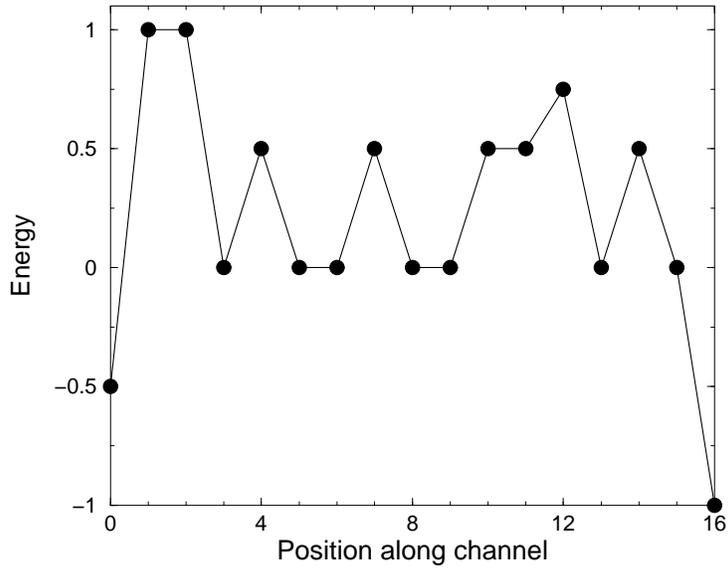}
\caption[]{
The dimensionless 
energy of each of the sites of a simple model for motion of a 
particle through a pore.  
}
\label{figIonE}
\end{figure}

\begin{figure}[t]
\includegraphics[width=.60\textwidth]{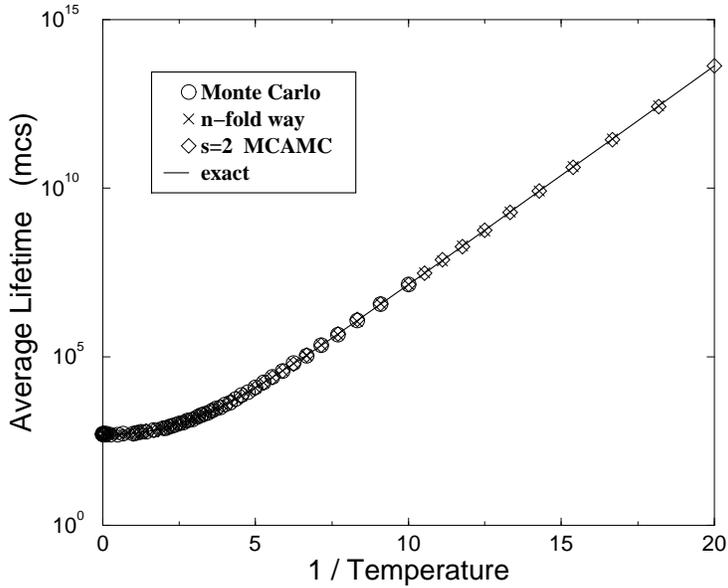}
\caption[]{
The average lifetime, in Monte Carlo steps (mcs), 
as a function of inverse temperature for 
particle motion in the energy arrangement of 
Fig.~\protect\ref{figIonE}.  
See Fig.~\protect\ref{figIonCPU} for an explanation of the plotting symbols.  
The solid line is the exact result using 
Eqs.~(\protect\ref{EIontau}), (\protect\ref{EIonPD15}), and 
(\protect\ref{EIonPDhi}).  
}
\label{figIonTau}
\end{figure}

At low temperatures, even the event-driven simulation can take a large 
amount of computer time.  This is principally due to two portions of the 
energy landscape, namely the lattice site pairs (5,6) and (8,9) 
in Fig.~\ref{figIonE}.  
In these situations it is advantageous to use more than one 
state in the transient subspace.  We will describe the case of two 
transient states, so a $s$$=$$2$ Monte Carlo with Absorbing Markov Chains (MCAMC) 
will be used whenever a walker steps into the pair (5,6) or 
the pair (8,9) in Fig.~\ref{figIonE}.  
At low temperature the probability of stepping out of this pair is 
exponentially small compared to the probability of stepping to the 
other site of the pair.  Consequently, the walker performs many steps 
within the pair before it eventually steps out of the pair of lattice sites.  
We have chosen the energies of these lattice pairs to make the $s$$=$$2$ 
MCAMC algorithm easy to describe.  The general case can easily be 
constructed using the absorbing Markov chain methodology of 
Sec.~3.0. 
In our case, the sites of the pair (5,6) or (8,9) have the same energy 
(chosen to be zero). 
Also, the bordering lattice sites 4 and 7 
(or 7 and 10) have the same energy.  
We will use the $s$$=$$1$ MCAMC, the event-driven simulation of 
Sec.~4.2, 
unless the walker has just stepped onto a lattice site pair (5,6) or (8,9).  
Let $x$ be the probability of stepping from 
site 5 to 4 in one time step, so 
\begin{equation}
\label{Es2x}
x=
{1\over 2} 
{ {\exp(-E_4/k_{\rm B}T)} \over
{\exp(-E_5/k_{\rm B}T)+\exp(-E_4/k_{\rm B}T)} }
\; .  
\end{equation}
The probabilities of stepping from 6 to 7 or from 8 to 7 or from 9 to 10 in 
one time step are also equal to $x$.  
The probability of stepping from 5 to 6 in one time step is ${1\over 4}$, 
which 
is also the probability of stepping from 6 to 5, 8 to 9, or 9 to 8.  

The Markov matrix for this $s$$=$$2$ transition is
\begin{equation}
{\bf M}_{\rm ion\ 2} = 
\pmatrix{
1 & 0 & 0 & 0 \cr
0 & 1 & 0 & 0 \cr
x & 0 & 1-x-{1\over 4} & {1\over 4} \cr
0 & x & {1\over 4} & 1-x-{1\over 4} \cr
}
\; .
\end{equation}
In our case we have
\begin{equation}
{\bf T}_2 
\pmatrix{1\cr 1\cr} 
=
\pmatrix{
1-x-{1\over 4} & {1\over 4} \cr
{1\over 4} & 1-x-{1\over 4} \cr
}
\pmatrix{1\cr 1\cr}
=
\left[1-x\right] 
\pmatrix{1\cr 1\cr}
\; .
\end{equation}
Thus the time $m$ to exit from the pair of lattice sites is given by 
\begin{equation}
\label{Es2m}
m = \left\lfloor {{\ln({\bar r})}\over{\ln(1-x)}} \right\rfloor + 1
\; .
\end{equation}
To calculate the probability of exiting to either side from a 
lattice site pair, for the normalization use 
\begin{equation}
{\vec v}_{\rm I}^{\rm T} {\bf T}_2^{m-1} \pmatrix{1\cr 1\cr} = 
(1-x)^{m-1}
\; .
\end{equation}
Also use the spectral decomposition of ${\bf T}_2$, which is
\begin{equation}
{\bf T}^{m-1}_2 = 
{1\over 2} (1-x)^{m-1} 
\pmatrix{1 \cr 1\cr} \pmatrix{1 & 1\cr} 
+
{1\over 2} 
\left({1\over 2}-x\right)^{m-1}
\pmatrix{1\cr -1\cr}
\pmatrix{1 & -1 \cr}
\; .
\end{equation}
This gives from Eq.~(\ref{EAMC5}) 
that the probability of exiting the lattice pair back to the same 
lattice point the walker entered from is
\begin{equation}
\label{Es2back}
p_{\rm exit\ back} = 
{1\over 2} \left[1+\left({{{1\over 2}-x}\over{1-x}}\right)^{m-1}\right]
\; ,
\end{equation}
and the probability of exiting to the lattice site opposite to the site 
it entered the lattice pair from is
\begin{equation}
\label{Es2other}
p_{\rm exit\ other} = 
{1\over 2} \left[1-\left({{{1\over 2}-x}\over{1-x}}\right)^{m-1}\right]
\; .
\end{equation}

The algorithm to use $s$$=$$2$ MCAMC for these two lattice pairs is:
\begin{itemize}
\item Place the walker at lattice site~0.
\item Set the time clock to $t$$=$$0$.
\item Perform the following steps until the walker is absorbed at lattice 
site~16.  
\begin{itemize} 
\item[$\circ$] Calculate two random numbers, ${\tilde r}$ and ${\bar r}$.  
\item[$\circ$] 
If the walker has not just stepped onto lattice site 5, 6, 8, or 9 
then use the rejection-free ($s$$=$$1$ MCAMC) algorithm 
described 
in Sec.~4.2: 
\begin{itemize}
\item[$\Box$] Calculate (or look up from tables) the two 
probabilities $p_{\rm right}$ and $p_{\rm left}$.  
\item[$\Box$] Calculate the time $m$ to exit from the current state 
$m$$=$$\left\lfloor 
{{\ln({\bar r})}\over{\ln(1-p_{\rm left}-p_{\rm right})}} 
\right\rfloor + 1$. 
\item[$\Box$] Move the walker to the right if 
${\tilde r}$$\le$${{p_{\rm right}}\over{p_{\rm right}+p_{\rm left}}}$, 
otherwise move the walker to the left.  
\end{itemize}
\item[$\circ$] If the walker has just stepped onto lattice site 
5, 6, 8, or 9 then use $s$$=$$2$ MCAMC: 
\begin{itemize}
\item[$\Box$] Calculate (or look up from tables) 
the probability $x$ from Eq.~(\ref{Es2x}).  
\item[$\Box$] Calculate the time $m$ to exit the 
pair of lattice points using Eq.~(\ref{Es2m}).  
\item[$\Box$] Calculate (or look up from tables) 
$p_{\rm exit\ back}$ from Eq.~(\ref{Es2back}).  
\item[$\Box$] If $p_{\rm exit\ back}$$<$${\bar r}$ then move the 
walker to the site it entered the lattice pair (5,6) or (8,9) 
from.  Otherwise move the walker to the other exit lattice site 
of the pair.  
\end{itemize}
\item[$\circ$] Increment the time to $t$$+$$m$.  
\end{itemize}
\end{itemize}

Using this $s$$=$$2$ MCAMC algorithm at $T$$=$$0.05$ 
for $K$$=$$1000$ escapes takes about 
0.35 minutes of CPU time.  
This is 
about 5000 times less than the $s$$=$$1$ MCAMC ($n$-fold way) algorithm, 
and about $10^{10}$ times faster than the traditional Monte Carlo algorithm!  

\begin{figure}[t]
\includegraphics[width=.40\textwidth]{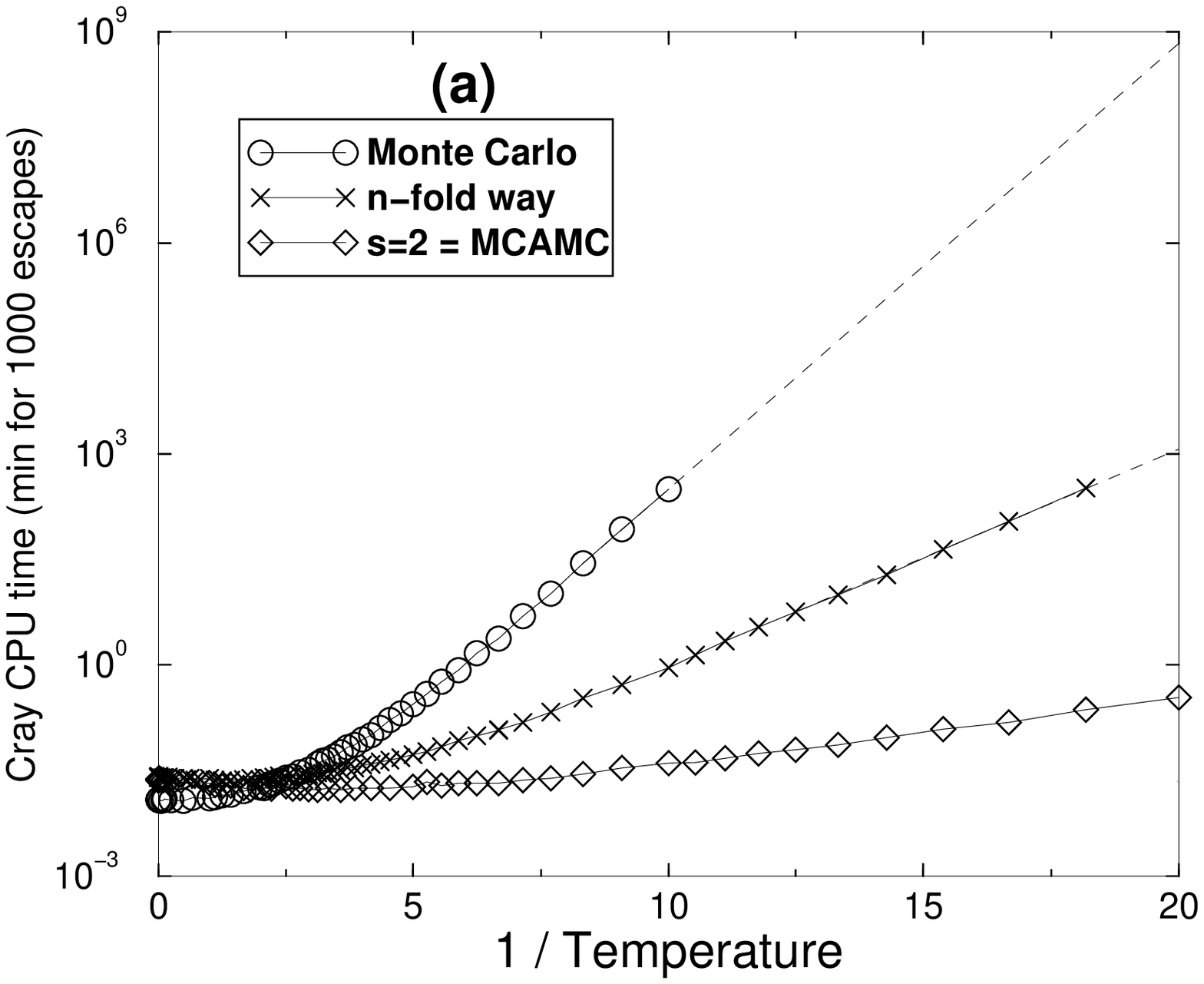}
\includegraphics[width=.40\textwidth]{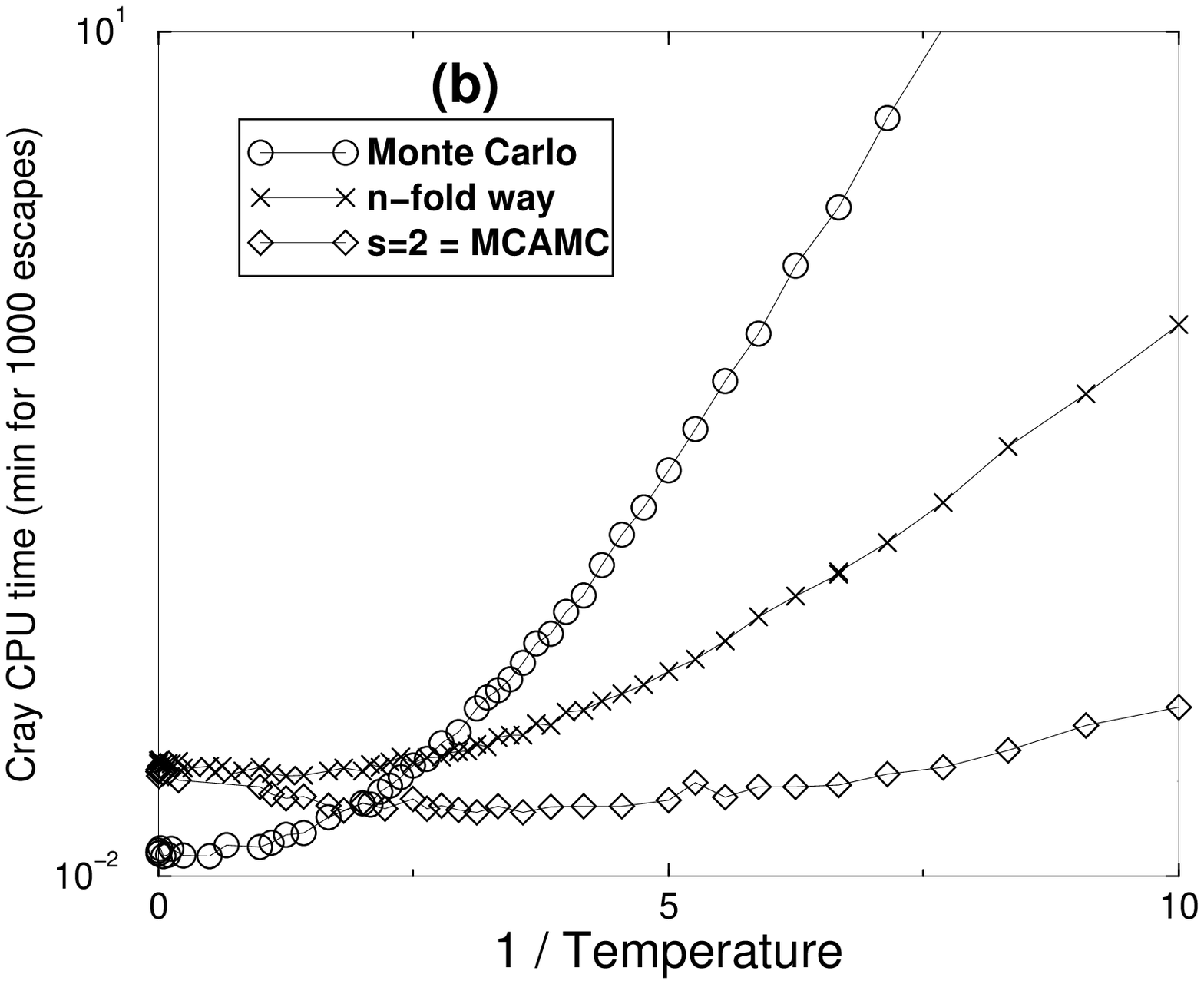}
\caption[]{
The amount of Cray CPU time (in minutes) required to 
perform $K$$=$$1000$ escapes as a function of inverse temperature.  
All simulations obtain the same average lifetime (within statistical errors) 
as shown in Fig.~\protect\ref{figIonTau}.  
The symbols correspond to 
standard Monte Carlo ($\circ$), 
rejection free or $n$-fold way or event driven or $s$$=$$1$ MCAMC ($\times$), 
and $s$$=$$2$ MCAMC for the lattice pairs (5,6) and (8,9) 
in Fig.~\protect\ref{figIonE} 
and 
$s$$=$$1$ MCAMC otherwise ($\diamond$).  
(a) The entire temperature range simulated, with solid lines joining the 
data points and the dashed lines extrapolated timings assuming that at low 
temperatures the CPU time required is an exponential in the 
inverse temperature.  
(b) The high-temperature portion of the same figure, showing where 
the algorithms cross over in speed.  
}
\label{figIonCPU}
\end{figure}

The timings for all three algorithms are shown in Fig.~\ref{figIonCPU}.  
The timing for the standard Monte Carlo algorithm is proportional to the 
average lifetime shown in Fig.~\ref{figIonTau}.  
Figure~\ref{figIonCPU}(b) shows the crossovers between the timings for 
the various algorithms.  The $s$$=$$2$ MCAMC turns out to be always 
at least as fast as the $s$$=$$1$ MCAMC.  This is not the usual case, and 
is due to the very simple form for ${\bf T}_2$.  
The standard Monte Carlo is faster at high temperatures, with the 
$s$$=$$2$ MCAMC being faster at temperatures below about $T$$=$${1\over 2}$.  

\subsection*{4.4~Ion Channel Projective Dynamics}
\label{Sec-Part4PD}

This simple model can actually be solved analytically by noting that the 
transient matrix is tridiagonal and has the form
\begin{equation}
{\bf T}_{16} = 
\pmatrix{
1-s_{15}-g_{15} & s_{15} & 0 & 0 & 0 & 0 \cr
g_{14} & 1-s_{14}-g_{14} & s_{14} & 0 & 0 & 0 \cr
0 & \ddots & \ddots & \ddots & 0 & 0 \cr
0 & 0 & g_2 & 1-s_2-g_2 & s_2 & 0 \cr
0 & 0 & 0 & g_1 & 1-s_1-g_1 & s_1 \cr
0 & 0 & 0 & 0 & g_0 & 1-g_0 \cr
}
\end{equation}
where $g_i$ is the single-time-step transition probability from 
$i$ to $i$$+$$1$ (the `growning probability') and 
$s_i$ is the single-time-step transition probability from 
$i$ to $i$$-$$1$ (the shrinking probability).  
In other words, we have changed notation slightly and 
$g_i$ are the appropriate $p_{\rm right}$ and 
$s_i$ are the appropriate $p_{\rm left}$.  
The walker starts in state $0$, so the initial vector is 
${\vec v}_{\rm I}^{\rm T} = 
\pmatrix{0 & 0 & \cdots & 0 & 1 }$.  
The recurrent matrix is given by 
\begin{equation}
{\bf R}_{16} = 
\pmatrix{
g_{15} \cr
0 \cr
\vdots \cr
0 \cr
}
\end{equation}
since the walker is absorbed when it steps onto state $16$ from state $15$.  

Introduce the average time $h(i)$ spent in state $i$ during this 
absorbing Markov process.  
Then the average lifetime is the sum of these residence times, 
\begin{equation}
\label{EIontau}
\langle \tau \rangle = \sum_{i=0}^{15} h(i)
\; .
\end{equation}
An iterative solution for the $h(i)$ terms can be found by considering the 
number of times during $K$ escapes that a walker passes an imaginary 
midpoint between state $i$ and $i$$-$$1$.  Since there are $K$ escapes and 
each escape must pass through each mid-point, there will be $K$ more 
walks to lattice site $i$ than to lattice site $i$$-$$1$.  
During $K$ walks the average time spent in state $i$ is $h(i)\>K$.  
Thus
\begin{equation}
g_{i-1} h(i-1) K = K + s_i h(i) K
\; .
\end{equation}
For site $16$ there is no shrinking probability and 
$g_{15} h(15) K = K$ or 
\begin{equation}
\label{EIonPD15}
h(15) = {1\over{g_{15}}}
\; .
\end{equation}
Solving for the other $h(i)$ values gives 
\begin{equation}
\label{EIonPDhi}
h(i-1) = {{1 + s_i h(i)}\over{g_{i-1}}}
\; .
\end{equation}
Thus we have a closed form for the lifetime for this simple process.  
This closed form is shown as the solid line in Fig.~\ref{figIonTau}.  
Note that the time unit is clock ticks, 
i.e.\ mcs (Monte Carlo trial steps).  

The model we used for the ion channel has only one entrance of the 
ion (from the left).  
A more realistic ion channel model has the possibility of 
entering and exiting from both ends of the channel \cite{BARC93}.  
For our simple 
model this only means that the state with no ions in the channel 
needs to be included together with the possibility of 
entering either end of the channel from this state and being 
absorbed by exiting from either end.  Again, the absorbing Markov 
chain is tridiagonal.  It now has two absorbing states.  
In addition to $\tau$, other quantities of interest include the 
average time to exit one end, given that it entered from a particular 
end.  These times can again be calculated using the 
methodology of absorbing Markov chains.  

In the sections below, 
we will utilize this formulation in the projective dynamics method for the 
escape of more complicated models from the 
metastable state \cite{GUNT83}.  In the majority of situations there is 
no analytic expression, but the same formulation can be used, 
see Sec.~5.4.  


\section*{5.0~Dynamic Ising Simulations}
\label{Sec-Ising}

The simple model for ion channels introduced above illustrates many of the 
advanced Monte Carlo methods, but it is too simple to completely 
illustrate their use.  We now describe 
advanced dynamic Monte Carlo simulations 
for a more complicated but still simple 
model, the square-lattice Ising ferromagnet 
described in Sec.~2.0. 
This model has 
$N$ Ising spins $\sigma$ 
which can be either up ($+1$) or down ($-1$) at each lattice site.  
We will consider the model with periodic boundary conditions on an 
$L$$\times$$L$ lattice, so $N$$=$$L^2$.  
We again are interested in the average lifetime 
$\langle\tau\rangle$ of the 
metastable state.  
The time units that correspond to physical time 
are Monte Carlo steps per spin, MCSS, as 
described in Sec.~2.0. 
The system starts with all spins up, and has an applied 
magnetic field $H$ which is down ($H$$\le$$0$).  
We consider only ferromagnetic 
nearest-neighbor interactions of strength $J$.  
The energy is 
given by Eq.~(\ref{EHIsing}).  
The total magnetization of the system 
is 
\begin{equation}
\label{EMI} 
M=\sum_i^{L^2}\sigma_{i=1}
\; .  
\end{equation}
This model has a critical temperature 
$k_{\rm B}T_{\rm c}$$=$
${2 J}$$/$${\ln(1+\sqrt{2})}$$\approx$$2.269\cdots$$J$.  
We want to measure the average lifetime $\langle\tau\rangle$ 
for $T$$<$$T_{\rm c}$ for the 
system to progress from the initial state 
with all spins up to a state with an 
equal number of spins up and down, i.e.\ to $M$$=$$0$.  
In this way we can study the dynamics of the model 
below the critical temperature, where 
the metastable decay of the model is dominated by 
nucleation and growth \cite{BIND73,BIND74,STAU82}.  

Even for this simple model the understanding of 
$\langle\tau\rangle$ is complicated.  Two recent reviews of 
nucleation and growth analysis of metastable decay in the Ising model 
are presented in 
\cite{RIKV94B,RIKV98}.  
A review of earlier works on metastable decay is given in 
Ref.~\cite{BIND76}.  
More complicated Ising models, 
including 
other boundary conditions 
\cite{RICH97A}, 
heterogeneous nucleation 
\cite{RICH97B}, 
and an approximate demagnetizing field 
\cite{RICH96} 
have been studied as well.  
The Ising model with periodic boundary conditions 
has four length scales \cite{RIKV94A,RIKV94B}.  
The lattice spacing $a$ and the lattice size $L$ 
are two of them.  Another length scale is the radius of the 
critical droplet, $R_{\rm c}(T,H)$, which is a function of temperature and 
applied field.  Droplets with radii smaller than $R_{\rm c}$ 
are subcritical and will most 
likely shrink, while droplets with radii larger than $R_{\rm c}$ 
are supercritical and will 
most likely grow.  The remaining length scale has been called 
$R_0(T,H)$ \cite{RIKV94A}, and it also depends on $T$ and the 
strength of the applied field.  
This can be regarded as the average size to which a supercritical droplet will 
grow before it encounters another supercritical droplet.  
Note that $R_0$$>$$R_{\rm c}$ always.  

\begin{figure}[t]
\includegraphics[width=.60\textwidth]{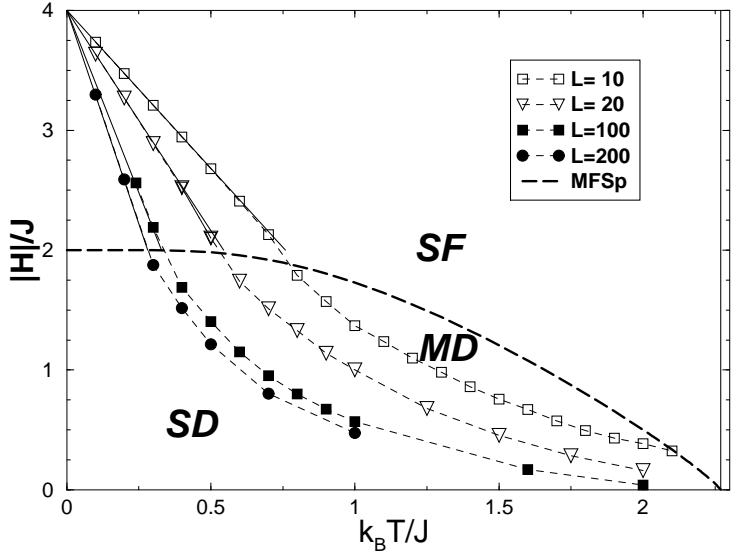}
\caption[]{
The cross-over `phase diagram for metastable decay' in the 
square-lattice Ising ferromagnet is shown.   
The strong-field (SF) regime, the multi-droplet (MD) regime, and 
the single-droplet (SD) regime are shown.  
The heavy dashed line separates the SF and MD regimes and is 
estimated using $R_{\rm c}$$=$$a/2$.  
The cross-overs between the SD and MD regimes depend on 
the system size and are shown for $L$$=$$10$, $20$, $100$, and $200$.  
The solid lines starting at $|H|$$=$$4J$ and $T$$=$$0$ are the 
single overturned spin cross-over between the SF and SD 
regimes, as described in the text.  
The vertical solid line shows the location of the critical 
temperature, $T_{\rm c}$.  
}
\label{figIsingCRX}
\end{figure}

These four length scales give rise to different types of decay 
mechanisms for the metastable state, depending on their 
inter-relationship.  
Only if $R_{\rm c}$ and $R_0$ are larger than $a$ is the droplet 
nucleation picture applicable, so when $R_c$$<$$a$ 
the system is in a strong-field (SF) regime where there is no 
long-lived metastable state.  
For weaker fields and for large lattices, 
$a$$\ll$$R_{\rm c}$$<$$R_0$$\ll$$L$.  Then many supercritical droplets 
form during the decay to $M$$=$$0$.  This is called the 
multi-droplet (MD) regime, and it is well described by the well-known 
Kolmogorov, Johnson-Mehl, Avrami theory \cite{AVRAETAL,RAMOS99}.  
For smaller lattices, where 
$a$$<$$R_{\rm c}$$\ll$$L$$\ll$$R_0$, only one supercritical droplet 
forms and grows to take the system to $M$$=$$0$, so this regime 
is called the single-droplet (SD) regime.  
For the square-lattice Ising model the crossovers between these various 
regimes are shown in Fig.~\ref{figIsingCRX} for 
different values of $L$.  
Another regime, where $L$$<$$R_{\rm c}$ is called the 
coexistence regime.   It is not shown in the figure.  

The cross-overs between the different decay regimes and how 
they depend on $H$ and $T$ are 
of central importance \cite{RIKV94A,RIKV94B}.  
A conservative estimate for the cross-over between the SF and 
MD regime is where $R_{\rm c}$$=$$a/2$.  This is shown as the 
heavy dashed curve in Fig.~\ref{figIsingCRX}, and is 
independent of $L$.  The cross-over between the 
MD and SD regime can be estimated by noting that the 
standard deviation of the lifetime, 
$\Delta\tau$$=$$\sqrt{\langle\tau^2\rangle-\langle\tau\rangle^2}$, 
approximately equals $\langle\tau\rangle$ 
in the SD regime and is much smaller than $\tau$ in the MD regime.  
One consequence of this is that the measurement of $\langle\tau\rangle$ is 
self-averaging in the MD regime, but is very far from self-averaging in 
the SD regime.  
Hence, to measure $\langle\tau\rangle$ in the MD regime only a few 
realizations are required, and the $K$$=$$1000$ escapes used 
here are more than adequate.  Conversely, in the SD regime the 
$K$$=$$1000$ escapes we average over are barely sufficient since the 
distribution of $\tau$ has an extremely long tail.  
The estimate for the cross-over between the MD and SD regimes, 
called the dynamic spinodal \cite{RIKV94A,TOMI92}, is given by 
the point where $\Delta\tau$$=$$\langle\tau\rangle/2$.  
The location of the dynamic spinodal 
depends on $L$, and is shown for 
various values of $L$ in Fig.~\ref{figIsingCRX}.  

For low temperatures and $|H|$$>$$2J$, a single overturned spin 
is the nucleating droplet 
\cite{NEVE91} 
with an energy barrier given by 
$\Gamma(H,J)$$=$$8J$$-$$2|H|$.  
The dynamic spinodal (DSp) for the random-update dynamic in 
Sec.~5.1 
can be found \cite{LEE95} by 
equating the average growth time to the average nucleation time 
to give 
\begin{equation}
\label{ElowTDSp}
{ {(4-H_{\rm DSp})} \over {k_{\rm B} T} } = 
{3\over 2} \ln(L) -0.82 
\; ,
\end{equation}
where the last term is the only fitting parameter and was 
chosen to provide agreement for values of $L$ between $8$ 
and $240$.  
The factor of ${3\over 2}$ changes if the dynamic changes, for 
example it is $1$ for sequential updates \cite{LEE95}.  
For a fixed field the dynamic spinodal approaches the 
$T$$=$$0$ axis as 
\begin{equation}
\label{ElowTDSp2}
k_{\rm B} T_{\rm DSp} 
= 
{
{4-|H|}
\over
{ {3\over 2} \ln(L) -0.82 } }
\; , 
\end{equation}
and without adequate finite-size scaling 
such a slow approach toward $T$$=$$0$ as a function of 
$L$ could easily be 
misinterpreted as a phase transition.  

For low temperatures, as seen in Fig.~\ref{figIsingCRX}, 
the single-droplet regime is dominant.  
However for strong fields the 
discreteness of the lattice becomes important and the 
nucleating droplets are no longer circular.  
For arbitrarily low temperatures and $|H|$$<$$2J$ with a fixed 
lattice size, the metastable decay 
is dominated by the SD regime, and 
the predicted lifetime is given by \cite{NEVE91} 
\begin{equation}
\label{ElowT}
k_{\rm B} T \ln(\langle\tau_{\rm low}\rangle) 
= \Gamma(H,J) = 8J\ell_c -2|H|(\ell_c^2-\ell_c+1) 
\; , 
\end{equation}
where 
\begin{equation}
\label{ElowTlc}
\ell_c = \left\lceil { {2J} \over {|H|} } \right\rceil
\; ,
\end{equation}
and $\lceil$$x$$\rceil$ denotes 
the smallest integer not less than $x$.  
The nucleating droplet is a cluster of overturned spins of 
the stable phase which is a $\ell_c$$\times$$(\ell_c$$-$$1)$ 
rectangle with one additional overturned spin on one of the 
longest sides of the rectangle \cite{NEVE91}.  
There is a non-trivial prefactor $A$ that enters the 
average lifetime \cite{NOVO97C} as 
\begin{equation}
\label{ElowTA}
\langle\tau_{\rm low\ SD}\rangle 
= A_{\rm SD} \exp\left[\Gamma(H,J)/k_{\rm B} T\right] 
\; .
\end{equation}
The prefactor $A_{\rm SD}$$=$$5/4$ for $\ell_c$$=$$1$ and 
$A_{\rm SD}$$=$$3/8$ for $\ell_c$$=$$2$ \cite{NOVO97C}.  
Closed forms for $A_{SD}$ for larger values of $\ell_c$ have 
not yet been calculated.  

The three limits, 
$L$$\rightarrow$$\infty$, 
$T$$\rightarrow$$0$, 
and 
$|H|$$\rightarrow$$0$ 
do not commute for the decay of the metastable state for 
the Ising model.  Taking first 
$L$$\rightarrow$$\infty$ places the system always in the 
MD regime for any finite temperature [see Fig.~\ref{figIsingCRX} 
and Eq.~(\ref{ElowTDSp2})].  
In this case the average lifetime is given by 
\cite{DEHG97} 
\begin{equation}
\label{ElowTMD}
\langle\tau_{\rm low\ MD}\rangle 
= A_{\rm MD} 
\exp\left[
{ {\Gamma(H,J)+(4J-2|H|) }\over{3 k_{\rm B} T} }
\right] 
\end{equation}
whenever $\ell_c$$>$$1$.  
The factor of $3$ 
(which is $d$$+$$1$ in $d$ dimensions)
in the denominator is due to the 
multi-droplet nature of the decay and is the same 
as that for the MD regime for circular droplets 
\cite{AVRAETAL,RAMOS99}.  
The additional term is due to the fact that at low temperatures 
for $|H|$$<$$2J$ the growth 
of supercritical droplets is dominated by nucleation events 
on the surface of these droplets 
\cite{DEHG97}.  

The advanced algorithms to be described below 
perform differently in the different decay regimes.  

\subsection*{5.1~Standard Monte Carlo for the Ising Model}
\label{Sec-IsingMC}

The Ising energies of Eq.~(\ref{EHIsing}) are sufficient to obtain 
the statics of the model, such as the specific heat or magnetic 
susceptibility.  However, this model does not have any dynamic 
since it is a classical model.  
A class of dynamics for this model can be derived using 
a Master equation approach \cite{GLAU63,BIND73,BIND74}.  
Martin in 1977 \cite{MART77} 
showed that it is possible to start with quantum spin $1\over 2$ 
particles each coupled to their own particular heat bath, and to 
obtain the dynamic described in Sec.~2.0. 

Note that there are an infinite number of possible dynamics that 
could obtain the correct statics with the energy of 
Eq.~(\ref{EHIsing}).  
For example, it is possible to use a Metropolis spin flip probability 
\cite{METR53,BIND86} 
given by Eq.~(\ref{EIMetrop}) rather 
than the Glauber spin flip probability Eq.~(\ref{EIGlauber}).  
It would also be possible to choose the spin in a particular order, 
perhaps sequentially.  
However, using one of these dynamics would lead to different values 
of $\langle\tau\rangle$ compared with the physically justified dynamic.  
For example, it has been shown \cite{RIKV94A,RIKV94B} 
that the power of the $H$ prefactor 
of the nucleation rate differs between the random update and 
sequential update dynamics, and the random update dynamic 
has an $H$  prefactor consistent with the value expected from 
field-theory arguments \cite{Lang60s,GUNT80}.  
Thus the physically motivated dynamic of Sec.~2.0 
should be used in cases 
where the dynamics is given a physical meaning.  
Furthermore, some of the advanced algorithms such as 
the $n$-fold way and MCAMC algorithms work only for 
the random update dynamic --- or would have to be modified 
significantly for the sequential update dynamic.  

The standard dynamic Monte Carlo for metastable decay of 
the Ising model is:
\begin{itemize}
\item Start the lattice with all $N$ spins up.  
\item Set the time $t$ to zero.  
\item Perform the following steps until the 
system reaches a magnetization $M$$=$$0$:  
\begin{itemize}
\item[$\circ$] Increment the time $t$ to $t$$+$$1$.
\item[$\circ$] Calculate two random numbers, ${\tilde r}$ and ${\bar r}$.
\item[$\circ$] Randomly choose one of the $N$ spins using ${\tilde r}$.  
\item[$\circ$] Calculate (or look up) the current energy $E_{\rm old}$ 
and the new energy $E_{\rm new}$ if the chosen spin were to flip.  
\item[$\circ$] Calculate (or look up from a table) the 
Glauber flip probability.  
\item[$\circ$] Use Eq.~(\ref{EIGlauber}) with ${\bar r}$ 
and either flip the spin or retain the old spin configuration.  
\end{itemize}
\end{itemize}
This algorithm is executed $K$ times, and the statistics for the 
mean lifetime are accumulated.  
A single algorithmic step is a Monte Carlo step (mcs), while 
the time that should be proportional to the physical time 
is a Monte Carlo step per spin (MCSS).  
The average lifetime, calculated for $K$$=$$1000$ escapes, is shown 
for two different fields in 
Fig.~\ref{figIsingTau1}(a) and 
Fig.~\ref{figIsingTau2}.  

\subsection*{5.2~$n$-fold Way for the Ising Model}
\label{Sec-Isingnfold}

The $n$-fold way algorithm introduced by Bortz, Kalos, and Lebowitz 
\cite{BORT75} 
uses spin classes for the Ising model.  
For the isotropic square-lattice Ising model with periodic boundary conditions 
in finite field, every spin belongs to one of $n$$=$$10$ different classes.  
These classes are organized by the spin orientation (up or down) and 
the number of nearest-neighbor spins that are up (which can be $0$, $1$, 
$2$, $3$, or $4$).  
The energy of every spin in a particular class is the same, so the 
probability of flipping any spin in a given class is the same.  
The $10$ spin classes are listed in Table~1.  
Note that for escape from the metastable state we start with all 
spins up so all spins are initially in class~1.  
When an up (down) spin flips, it changes its spin class by 
$+5$ ($-5$).  Furthermore, when an up (down) spin flips, all four of 
its nearest-neighbor spins change their spin classes by 
$+1$ ($-1$).  Also shown in Table~1 
are the new and old local energies 
of the spins --- remember $J$$>$$0$ while $H$$<$$0$ for our 
escape from the metastable state.  

\begin{table}
\label{TableI} 
\begin{tabular}{|r|c|c|r|r|} 
\hline
\hline
Class  & Spin        & Number of       & $-E_{\rm old}$ & $-E_{\rm new}$ \\
Number & Orientation & nearest-neighbor &                &                \\
       &             & spins up        &                &                \\
\hline
\hline
 1 & $\uparrow$   & 4 &  $4J+H$ & $-4J-H$ \\
 2 & $\uparrow$   & 3 &  $2J+H$ & $-2J-H$ \\
 3 & $\uparrow$   & 2 &    $+H$ &    $-H$ \\
 4 & $\uparrow$   & 1 & $-2J+H$ &  $2J-H$ \\
 5 & $\uparrow$   & 0 & $-4J+H$ &  $4J-H$ \\
\hline
 6 & $\downarrow$   & 4 & $-4J-H$ &  $4J+H$ \\
 7 & $\downarrow$   & 3 & $-2J-H$ &  $2J+H$ \\
 8 & $\downarrow$   & 2 &    $-H$ &    $+H$ \\
 9 & $\downarrow$   & 1 &  $2J-H$ & $-2J+H$ \\
10 & $\downarrow$   & 0 &  $4J-H$ & $-4J+H$ \\
\hline
\hline
\end{tabular}
\caption{The $10$ classes of Ising spins for the square-lattice 
nearest-neighbor model.}
\end{table}

Let $n_i$ be the number of spins in class~$i$.  
One always has
\begin{equation}
\label{EsumN}
N = \sum_{i=1}^{10} n_i 
\; .
\end{equation}
Let $p_i$ be the probability of flipping a spin in class $i$ 
using the dynamic such as the Glauber dynamic of 
Eq.~(\ref{EIGlauber}), 
given that the spin was chosen during the random choice 
portion of the dynamic.  
Then the probability in one time step 
of flipping any spin in class $i$ is $n_i p_i/N$ 
since $n_i/N$ is the probability of randomly choosing 
a spin in class $i$ and $p_i$ is the flip probability once a spin 
in class $i$ 
has been chosen.  
In order to exit from the current spin configuration, a spin in one of 
the $10$ classes must flip.  Consequently, the absorbing Markov 
chain has one transient state and $10$ absorbing states.  
In general, the $n$-fold way has $n$ classes for a particular discrete 
spin model, and there are $n$ absorbing states and one transient 
state, the current spin configuration.  

The original $n$-fold way algorithm was written in continuous time 
\cite{BORT75}.  
It was recast in a form for discrete time in 1995 \cite{NOVO95}.  
We will use the discrete-time version because it allows for a much easier 
generalization to more than one transient state.  
Introduce 
\begin{equation}
\label{EintroQ}
Q_i = {1\over N} \sum_{j=1}^i n_j p_j \quad\quad 1\le i\le n
\end{equation}
and define $Q_0$$=$$0$.  
The absorbing Markov matrix for exiting from the current spin configuration is 
\begin{equation}
\label{EAMCnfold}
{\bf M}_{n+1} = 
{1\over N}
\pmatrix{
N & 0 & 0 & 0 & 0 & 0 & 0 & 0 & 0 & 0 & 0 \cr
0 & N & 0 & 0 & 0 & 0 & 0 & 0 & 0 & 0 & 0 \cr
0 & 0 & N & 0 & 0 & 0 & 0 & 0 & 0 & 0 & 0 \cr
0 & 0 & 0 & N & 0 & 0 & 0 & 0 & 0 & 0 & 0 \cr
0 & 0 & 0 & 0 & N & 0 & 0 & 0 & 0 & 0 & 0 \cr
0 & 0 & 0 & 0 & 0 & N & 0 & 0 & 0 & 0 & 0 \cr
0 & 0 & 0 & 0 & 0 & 0 & N & 0 & 0 & 0 & 0 \cr
0 & 0 & 0 & 0 & 0 & 0 & 0 & N & 0 & 0 & 0 \cr
0 & 0 & 0 & 0 & 0 & 0 & 0 & 0 & N & 0 & 0 \cr
0 & 0 & 0 & 0 & 0 & 0 & 0 & 0 & 0 & N & 0 \cr
n_1 p_1 & n_2 p_2 & n_3 p_3 & n_4 p_4 & n_5 p_5 & n_6 p_6 & 
n_7 p_7 & n_8 p_8 & n_9 p_9 & n_{10} p_{10} & 
{\bf T}_{1,1} 
}
\end{equation}
where 
${\bf T}_{1,1}$$=$$N(1-Q_{10})$ is the 
transient matrix, here with a single element.  

The algorithm to perform the $n$-fold way for escape from the 
metastable state is:
\begin{itemize}
\item Start the lattice with all $N$ spins up.  
Set $n_1$$=$$N$ and $n_i$$=$$0$ for $i$$>$$1$ and 
also set $M$$=$$N$.  
\item Set the time $t$ to zero.  
\item Perform the following steps until the 
system reaches a magnetization $M$$=$$0$:  
\begin{itemize}
\item[$\circ$] Calculate three random numbers, 
${\tilde r}$, ${\bar r}$, and $r$. 
\item[$\circ$] Calculate the 10 values for $Q_i$ using 
Eq.~(\ref{EintroQ}).  
\item[$\circ$] 
Calculate the time $m$ (in units of Monte Carlo spin flip attempts) 
to exit from the current spin 
configuration (state) using 
\begin{equation}
\label{Einfoldm} 
m = 
\left\lfloor 
{{\ln({\bar r})}\over{\ln(1-Q_{10})}} \right\rfloor + 1
\; . 
\end{equation} 
\item[$\circ$] Using the random number ${\tilde r}$, 
find the spin class $k$ that satisfies 
\begin{equation}
\label{EnfoldR}
Q_{k-1} \le {\tilde r}Q_{10} < Q_k
\; .  
\end{equation}
\item[$\circ$] Use the random number $r$ to pick one of the 
$n_k$ spins from spin class $k$, and flip the chosen spin.  
\item[$\circ$] Change the number of spins in class $k$ to $n_k$$-$$1$.  
\item[$\circ$] If the flipped spin was up (down), change the 
number of spins in class $n_{k+5}$ ($n_{k-5}$) by $+1$.  
\item[$\circ$] If the flipped spin was up (down), 
for each of the four nearest-neighbor spins initially in class $j$ 
change the number of spins in class $j$ to be $n_j$$-$$1$ and 
the number of spins in class $n_{j+1}$ ($n_{j-1}$) to be 
$n_{j+1}$$+$$1$ ($n_{j-1}$$+$$1$).   
\item[$\circ$] Set the time $t$ to $t$$=$$t$$+$$m$. 
\end{itemize}
\end{itemize}

A number of remarks are in order.  
First, just as in the $n$-fold way for the simple ion channel, the 
time to exit the current state comes from 
using Eq.~(\ref{EAMC3}) with the Markov matrix of 
Eq.~(\ref{EAMCnfold}).  
Since there is only one transient state, the 
equation 
\begin{equation}
\label{Enfoldm0}
\left(1-Q_{10}\right)^{m} 
< {\bar r} 
\le \left(1-Q_{10}\right)^{m-1} 
\end{equation}
reduces to Eq.~(\ref{Einfoldm}).  
Second, 
the state to which the current spin configuration changes 
comes from 
using Eq.~(\ref{EAMC5}) with the Markov matrix of 
Eq.~(\ref{EAMCnfold}).  
Since there is only one transient state, the factors of 
${\bf T}^{m-1}$ in Eq.~(\ref{EAMC5}) 
are scalars and cancel, and the normalization 
in Eq.~(\ref{EAMC5}) becomes 
${\vec v}_{\rm I}^{\rm T} {\bf R} {\vec e}$$=$$Q_{10}$.  

To recover the original \cite{BORT75} continuous-time result, 
note that $Q_{10}$$<$$1$ always, and usually $Q_{10}$$\ll$$1$ 
for most spin configurations near a minimum of the 
free energy (i.e.\ near a metastable or stable state).  
Expanding Eq.~(\ref{Einfoldm}) gives 
\begin{eqnarray}
\label{Enfoldt}
m & = & 
\left\lfloor {\ln({\bar r})}
\left({1\over{\ln(1-Q_{10})}}\right) \right\rfloor + 1
\nonumber 
\\ 
 & \approx &
\left\lfloor \ln({\bar r})
\left[{ 
-{1\over Q_{10}} + {1\over 2} 
+{1\over {12}} Q_{10} 
+{1\over {24}} Q_{10}^2  
+{{19}\over {720}} Q_{10}^3 + \cdots   
}\right]
\right\rfloor + 1
\; . \\
\end{eqnarray}
Keeping only the first term in Eq.~(\ref{Enfoldt}) gives the time 
increment 
\begin{equation}
\label{EnfoldDt}
m \approx \Delta t = 
-{{\ln({\bar r})}\over{Q_{10}}} 
\; ,
\end{equation}
with the time units in Monte Carlo steps (mcs).  
To change the time units to physical time units (MCSS) 
for $\Delta t$, one needs to divide 
Eq.~(\ref{EnfoldDt}) by $N$.  
The time unit is trivial to 
change to MCSS in the continuum $n$-fold way since it just cancels the 
factor of $N$ in the definition for the 
$Q_i$, Eq.~(\ref{EintroQ}), and gives the result in 
Ref.~\cite{BORT75} (where the authors 
define their values of $Q_i$ without the $1/N$ normalization).  
However, no such fortuitous cancelation between $m$ and 
$N$ during a time unit change occurs in the discrete version of the 
$n$-fold way, Eq.~(\ref{Einfoldm}).  

There is an efficient way 
\cite{BORT75} to keep track 
of the classes of all the spins that uses four arrays. 
(See Ref.~\cite{BORT75} for the use 
of three arrays and the additional packing of information). 
The 
array {\tt NNCLS(10)} contains the number of spins $n_i$ in each
of the 10 classes. The two-dimensional array {\tt LOC($N$,10)} contains
the {\tt LOC\/}ation of the spins in the lattice for each of the 10 classes
of spins, where the $L$$\times$$L$ lattice is considered as a
one-dimensional array of spins. 
There is no particular order within
a class of spins, and only the {\tt NNCLS(i)} spins in 
{\tt LOC(j,i)} contain spin locations in current use. 
(Adjustable
partitions in a single array of length $N$ could be used for 
{\tt LOC\/} \cite{BORT75}, 
but in {\tt FORTRAN\/} a price is paid in terms of the time 
required to do 
the bookkeeping involved in changing the adjustable partitions after
every time step. Here {\tt C\/} and {\tt C++\/} programmers have an 
advantage.) 
Array {\tt LOOK($N$)} determines the index
of the first dimension in LOC for each of the $N$ spins, and the
array {\tt LOOC($N$)} is used to determine the class of a spin in
{\tt LOC}, given its position in the lattice. 
Hence ${\tt LOC(LOOK(i),LOOC(i))=i}$.

Suppose we have chosen class $k$ using 
$Q_{k-1}$$\le$${\tilde r}Q_{10}$$<$$Q_k$ 
and want to flip one of the $n_k$ spins 
in this class. We first calculate a uniform random integer
$j$ between 1 and {\tt NNCLS(k)} and find the spin location 
$i={\tt LOC(}j,k{\tt )}$. Then we need to update all four arrays for
the change in classes of the 
$i$th spin and its four nearest neighbors. This indexing provides a
fast method for changing the classes of spins in the $n$-fold way
algorithm \cite{BORT75,BKTERR}.  

\begin{figure}[t]
\includegraphics[width=.50\textwidth]{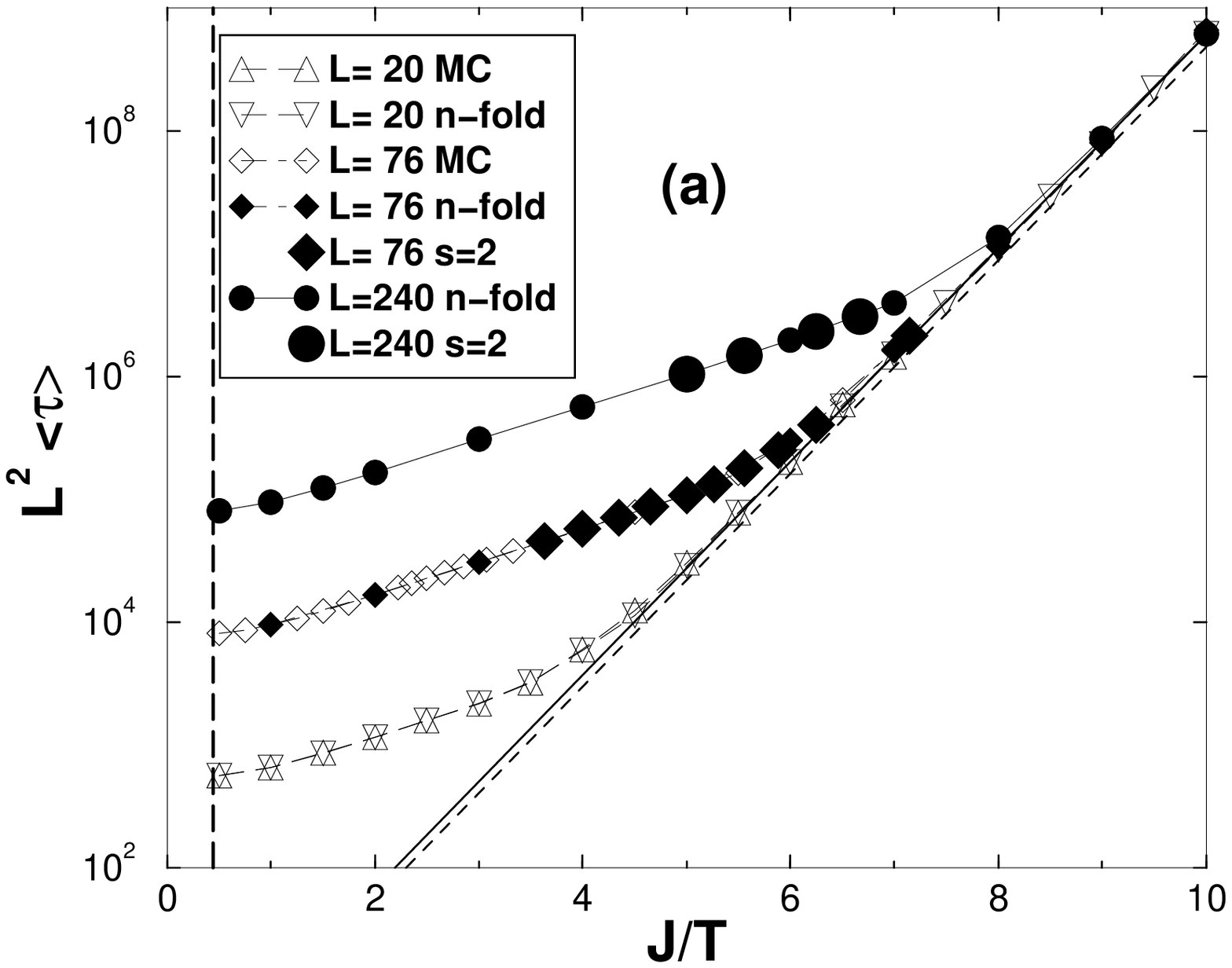}
\includegraphics[width=.50\textwidth]{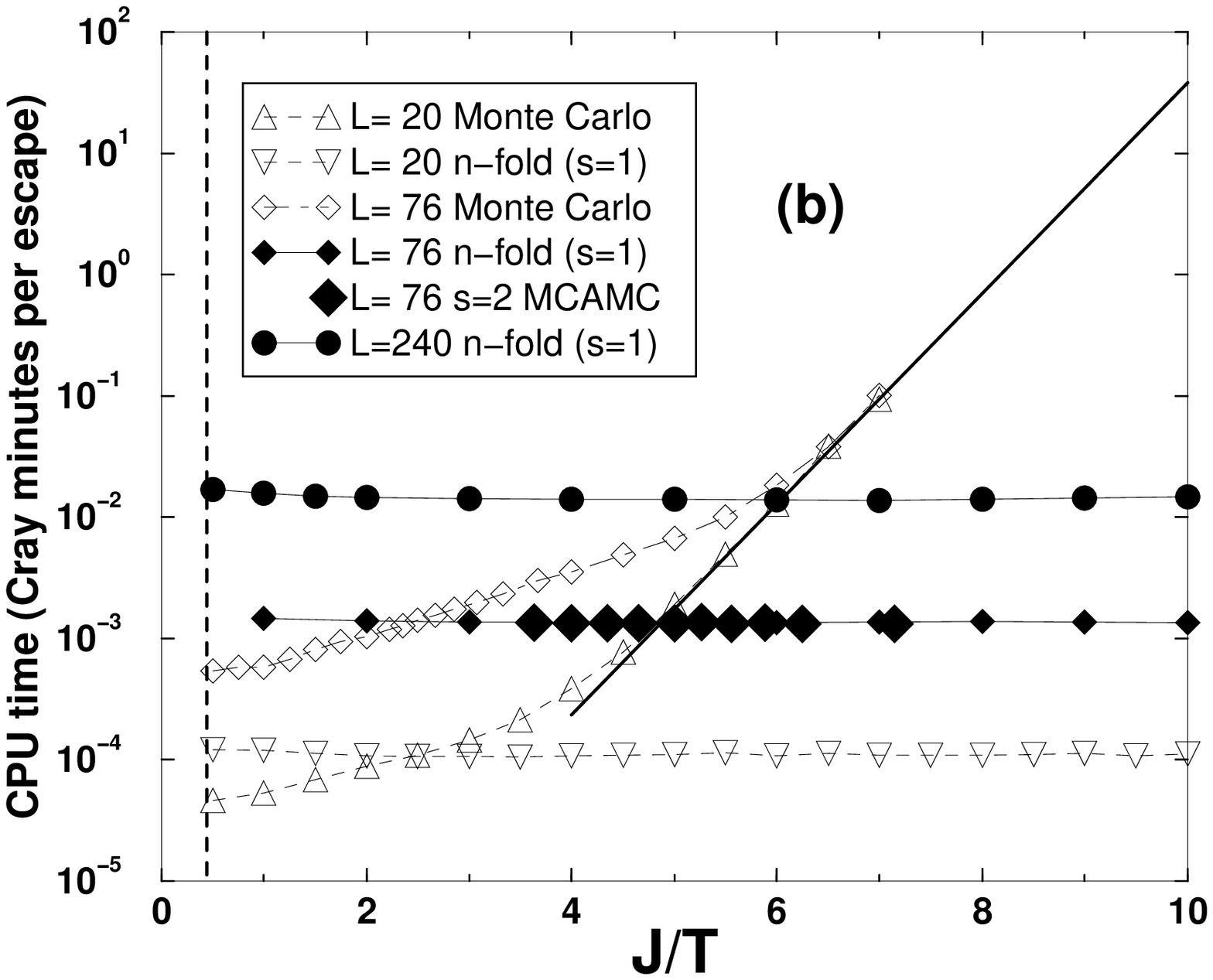}
\caption[]{
For $H$$=$$-3$$J$ with $L$$=$$20$, $L$$=$$76$, and $L$$=$$240$ 
data are shown for different dynamic Monte Carlo algorithms.  
(a) $L^2$ times 
the average lifetime, 
$L^2$$\langle\tau\rangle$, 
for the square-lattice Ising ferromagnet 
is shown as a function of inverse temperature, note the 
units of $L^2$ times MCSS which is 
Monte Carlo steps (mcs).  
The vertical dashed line is the critical temperature.  
The dashed line corresponds to 
the low-temperature prediction 
$\langle\tau\rangle$$=$$\exp(2J/k_{\rm B}T)$ 
\protect\cite{NEVE91}, 
while the solid line corresponds to 
$\langle\tau\rangle$$=$$(5/4)$$\exp(2J/k_{\rm B}T)$, 
which includes the exact low-temperature prefactor \protect\cite{NOVO97C}.  
Note that all algorithms produce the same lifetimes within statistical 
errors.  
In (b) the CRAY CPU time in minutes per escape is 
shown.  Note that for the $n$-fold way algorithm (rejection free, 
$s$$=$$1$ MCAMC) the CPU time per escape is almost flat.  This is 
because for $2J$$<$$|H|$$<$$4J$ the nucleating droplet is a 
single overturned spin \protect\cite{NEVE91}.  
The $s$$=$$2$ MCAMC algorithm, shown only for $L$$=$$76$, is also 
approximately flat.  
Conversely, the standard dynamic Monte Carlo 
algorithm requires a CPU time that grows exponentially with 
inverse temperature, 
shown as the heavy solid line.  
}
\label{figIsingTau1}
\end{figure}

\begin{figure}[t]
\includegraphics[width=.40\textwidth]{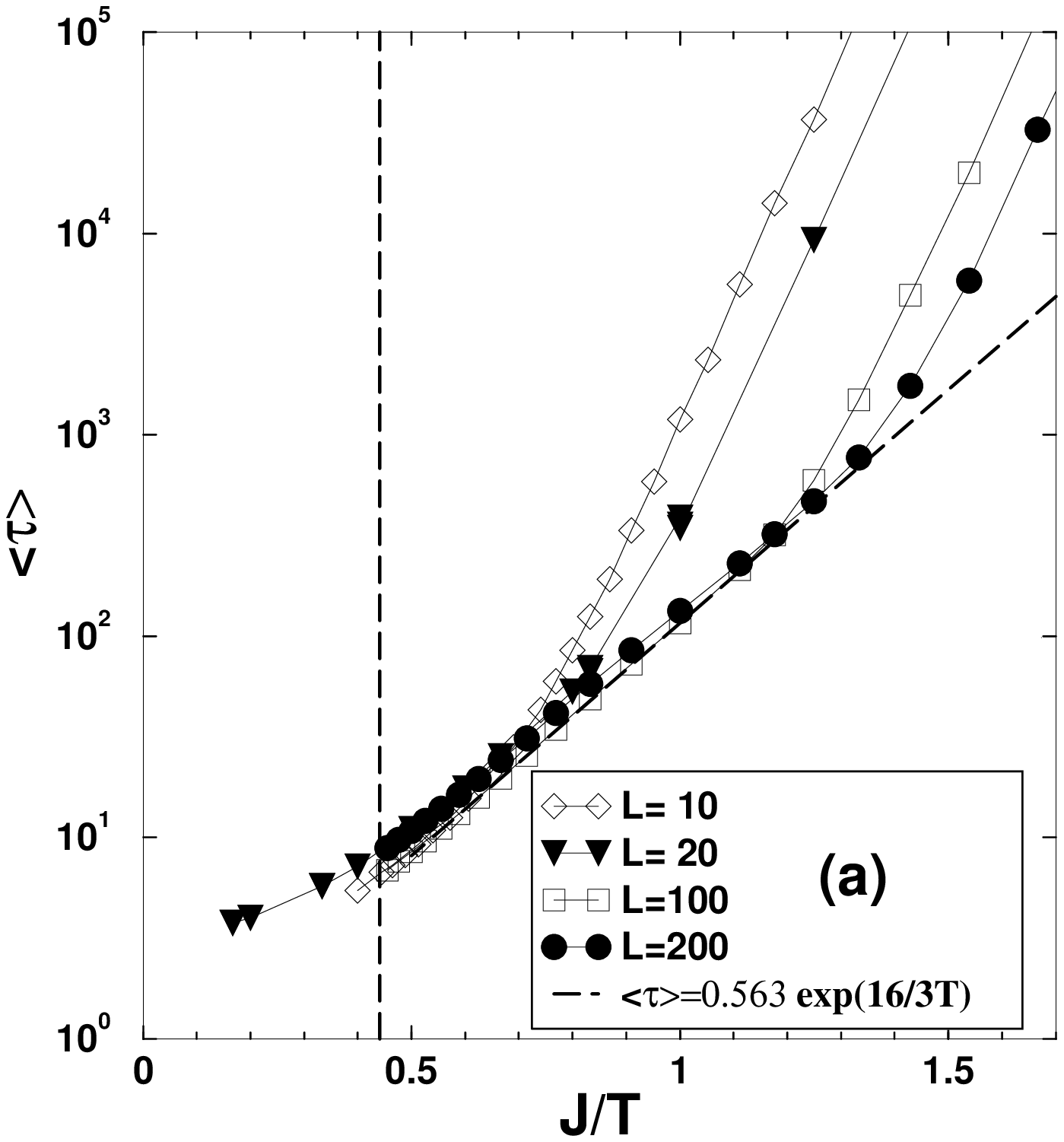}
\includegraphics[width=.40\textwidth]{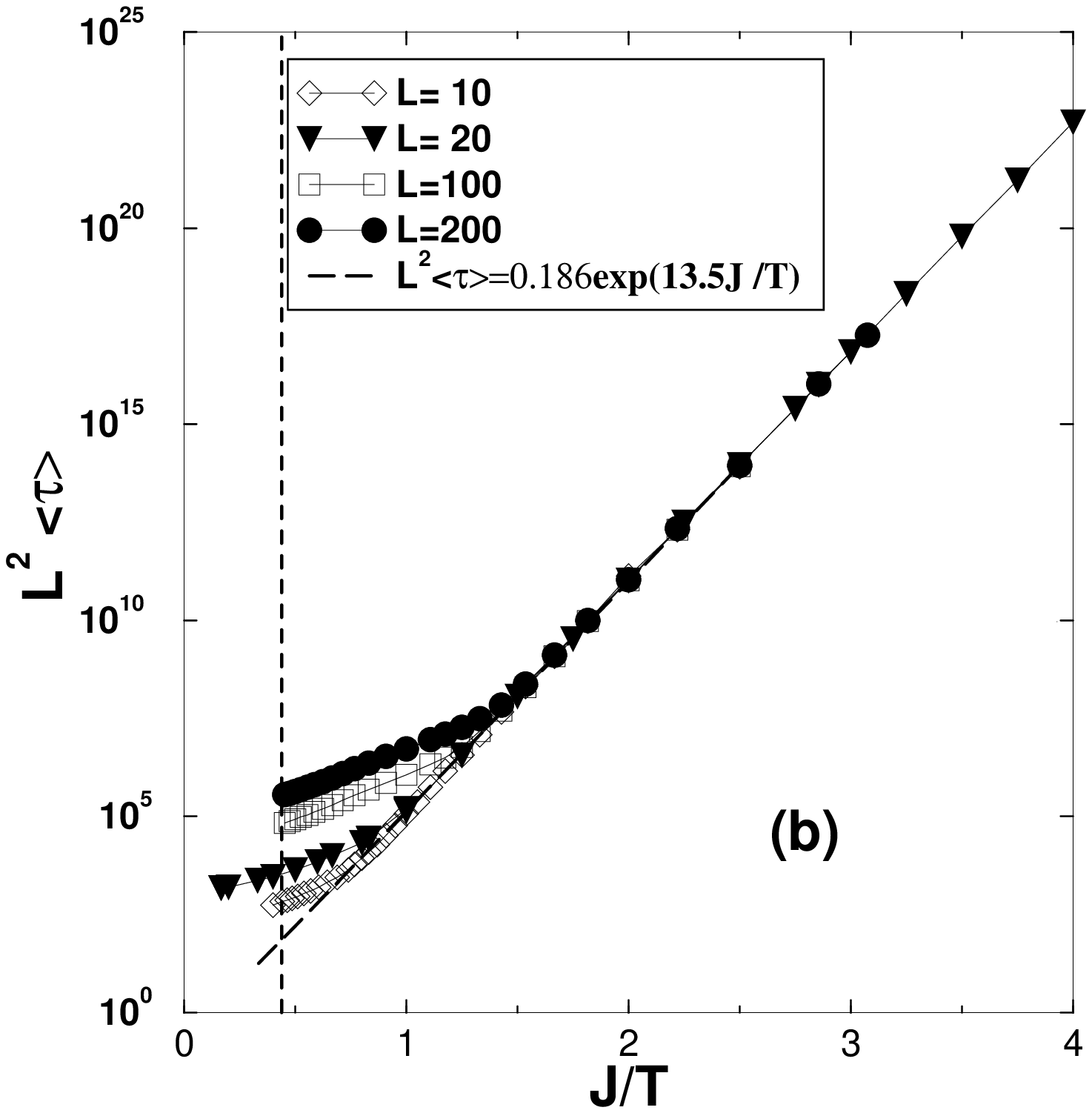}
\caption[]{
The average lifetime, $\langle\tau\rangle$, 
for the square-lattice Ising ferromagnet 
is shown as a function of inverse temperature for 
$H$$=$$-0.75$$J$ for different $L$.  
The same data are plotted 
in (a) and (b) in different time units.  The vertical dashed line is 
the critical temperature.  
In (a) the lifetime $\langle\tau\rangle$ 
is shown in units of the physical time, Monte Carlo Steps per Spin (MCSS).  
The dashed line corresponds to 
$\langle\tau\rangle$$=$$0.563\exp(16J/3T)$ 
while the light solid lines connect the data points.  
In (b) the lifetime $\langle\tau\rangle$ 
is shown in units of MCSS times $L^2$ (or Monte Carlo 
steps --- mcs).  The heavy dashed line corresponds to 
$\langle\tau\rangle$$=$$0.186$$L^{-2}$$\exp(13.5J/T)$ 
and the light solid lines connect the data points.  
In (a) the data points lie on top of each other in the SF and MD regime 
since the lifetime there is independent of $L$.  
In (b) the data points in the SD regime lie on top of each other since the 
lifetime there is inversely proportional to the volume.  
}
\label{figIsingTau2}
\end{figure}

\begin{figure}[t]
\includegraphics[width=.40\textwidth]{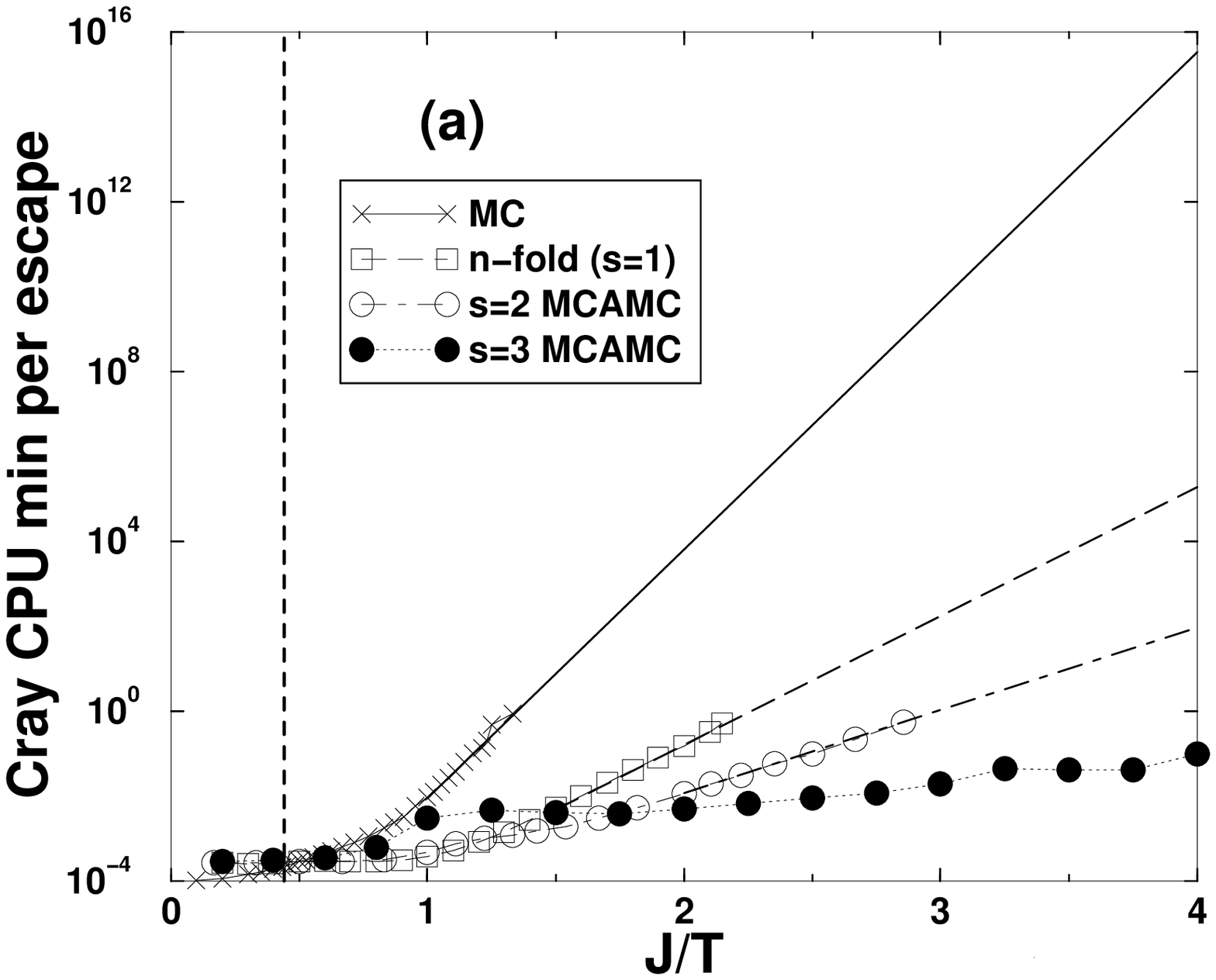}
\includegraphics[width=.40\textwidth]{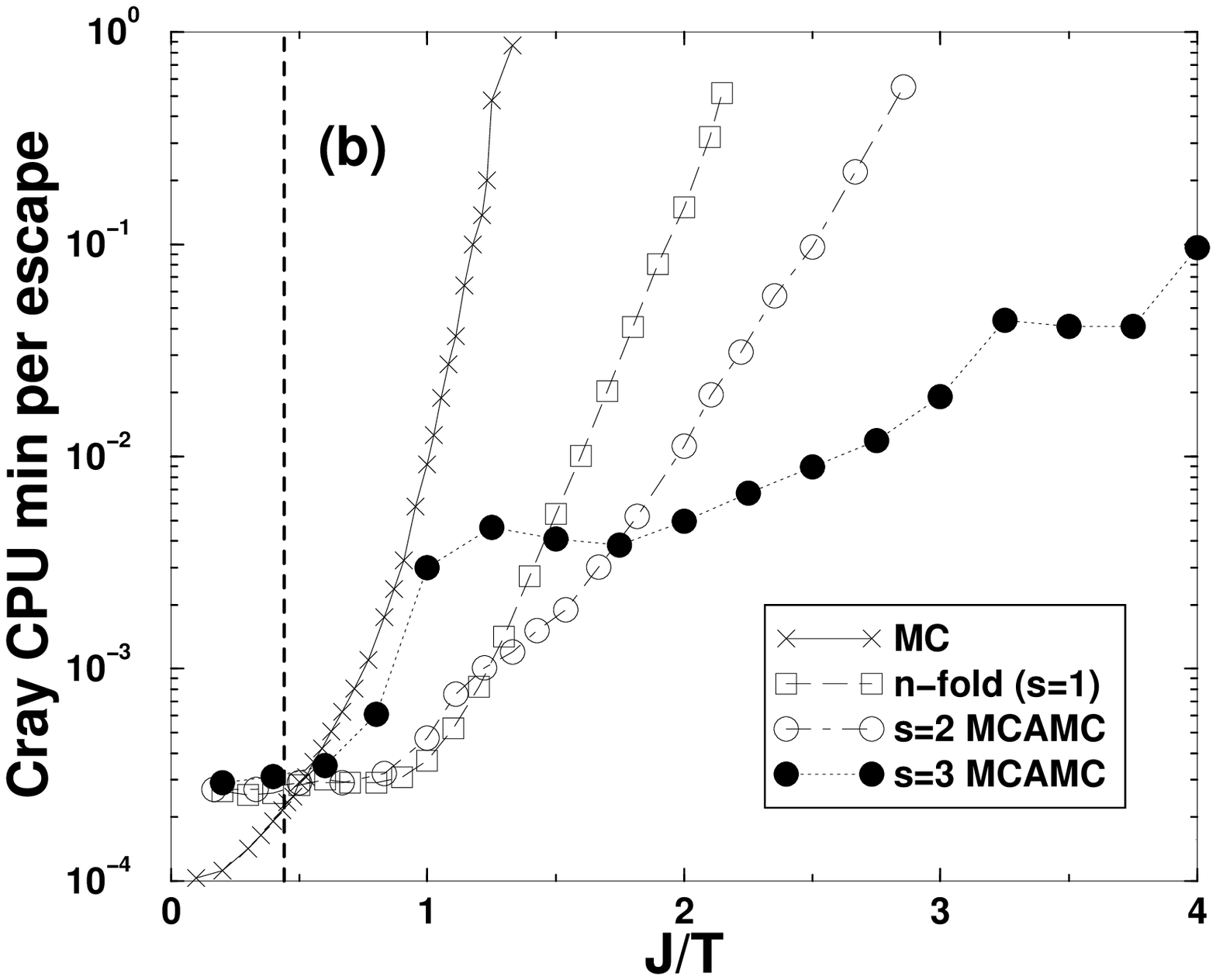}
\caption[]{
The CPU time required for the simulation of the 
$20$$\times$$20$ Ising ferromagnets presented in 
Fig.~\protect\ref{figIsingTau2} is shown 
as a function of inverse temperature.  
The symbols are for standard dynamic Monte Carlo ($\times$), 
$n$-fold way simulations ($\Box$), 
$s$$=$$2$ MCAMC ($\circ$), and 
$s$$=$$3$ MCAMC ($\bullet$).  
(a) Uses a fit to 
$A_{\rm fit} \exp\left[(\Gamma-\Gamma_0)/k_{\rm B}T\right]$
to estimate the low-temperature CPU time required 
for the slower algorithms.  
This fit is given by the corresponding heavy lines.  
The fit parameters are listed in Table~2. 
(b) Shows the region where simulations were 
actually performed.  This shows the cross-overs between 
the timings of the various algorithms.  
}
\label{figHItime1}
\end{figure}

The CPU time required for the data presented in 
Fig.~\ref{figIsingTau1}(a) is presented in 
Fig.~\ref{figIsingTau1}(b) 
and 
the CPU time required for the data presented in 
Fig.~\ref{figIsingTau2}
are shown in 
Fig.~\ref{figHItime1}.  
Note that the $n$-fold way algorithm is significantly faster 
at low temperatures (by orders of magnitude!) than the 
standard dynamic Monte Carlo algorithm.  
For $H$$=$$-3J$ the CPU time required for the $n$-fold way 
simulation is roughly independent of the temperature 
for any value of $L$.  This is 
because for $2J$$<$$|H|$$\le$$4J$ a single overturned spin 
is the nucleating droplet \cite{NEVE91}.  For $|H|$$<$$2J$, as in 
Fig.~\ref{figIsingTau2}, 
the CPU time required for the $n$-fold way simulation increases 
as the temperature is lowered since the nucleating droplet 
includes more than one overturned spin 
\cite{NEVE91}.  See Fig.~\ref{figHItime1}.  

\subsection*{5.3~MCAMC for the Ising Model}
\label{Sec-IsingMCAMC}

Although the $n$-fold way algorithm is extremely fast for 
$2J$$<$$|H|$, it is possible to further decrease the amount of CPU time 
required for a given simulation with $|H|$$<$$2J$ by using more 
states in the transient subspace.  In particular, at low 
temperatures where the energy of the 
nucleating droplet is known \cite{NEVE91} 
to be given by Eq.~(\ref{ElowT}), 
it is possible to use higher $s$ MCAMC only for the 
exit from the state with all spins up and use 
$s$$=$$1$ MCAMC ($n$-fold way) for the other moves.  
This is similar to the strategy used in the MCAMC for the 
ion-channel model of Sec.~4.3. 

For $s$$=$$2$ MCAMC the transient matrix is 
\begin{equation}
\label{Es2IT} 
{\bf T}_{s=2} = 
\pmatrix{
1-{{\left[p_6+4p_2+(N-5)p_1\right]}\over N} & {p_6\over N}  \cr
p_1                                         & 1-p_1         \cr
}
\end{equation}
and the recurrent matrix is 
\begin{equation}
\label{Es2IR} 
{\bf R}_{s=2} = 
{1\over N} 
\pmatrix{
 4p_2   & (N-5)p_1 \cr 
    0   &       0  \cr
}
\; .  
\end{equation}
The first transient state (in the lower right-hand corner of the matrix) 
is the initial state (all spins up), and the second state 
is all the $N$ spin configurations with one overturned spin.  The first 
absorbing state corresponds to all $2N$ spin configurations with two 
nearest-neighbor overturned spins, and the second 
to all $N(N$$-$$5)/2$ spin configurations with 
two overturned spins that are not nearest-neighbor spins.  
In the MCAMC algorithm, if the spin configuration has two or more overturned 
spins, the normal $n$-fold way algorithm is used.  If the 
system returns to the initial state (all spins up), the 
absorbing Markov chain part 
of the algorithm 
using Eqs.~(\ref{Es2IT}) and (\ref{Es2IR}) 
is performed.  

For $s$$=$$3$ the transient matrix for starting from the 
state with all spins up is 
\begin{equation}
\label{Es3IT} 
{\bf T}_{s=3} = 
{1 \over N} 
\pmatrix{
N-2p_7-6p_2-(N-8)p_1 & 2 p_7               & 0        \cr
4p_2                 & N-p_6-4p_2-(N-5)p_1 & p_6      \cr
0                    & Np_1                & N(1-p_1) \cr
}
\end{equation}
and the recurrent matrix is
\begin{equation}
\label{Es3IR} 
{\bf R}_{s=3} = 
{1 \over N} 
\pmatrix{
(N-8)p_1 & 2 p_2 & 4p_2 & 0        \cr
0        & 0     & 0    & (N-5)p_1 \cr
0        & 0     & 0    & 0        \cr
}
\; .
\end{equation}
Here the transient state added to the $s$$=$$2$ MCAMC is 
all $2N$ spin configurations with two overturned nearest-neighbor spins.  
The absorbing states correspond (left to right) to 
all the spin configurations with: 
1) two nearest-neighbor overturned spins and one not-connected 
overturned single spin; 
2) a linear chain of three overturned spins; 
3) an L-shaped arrangement with three overturned spins --- 
which corresponds to the nucleating droplet at low 
temperatures for $J$$<$$|H|$$<$$2J$ which corresponds to 
$\ell_c$$=$$2$ of Eq.~(\ref{ElowTlc}); 
4) two not-nearest-neighbor overturned spins.  

After exiting from the absorbing Markov Chain portion of 
the algorithm, the spin configuration on exit must be decided on 
since each of the absorbing states corresponds to a large 
number of spin configurations with equal absorption probabilities. 
The method used in the $n$-fold way algorithm to decide on 
which particular spin within a chosen spin class is generalized 
to accomplish this.  
For example, the first absorbing state 
for the $s$$=$$3$ MCAMC uses three random numbers, $r_i$, 
to construct the exiting spin configuration.  
First $r_1$ is used to pick uniformly one of the $N$ spins 
and this spin is flipped to down.  (With periodic boundary conditions this step is 
not strictly necessary, but is advisable to avoid any 
correlations between the random number and the algorithm 
\cite{DPLprl} --- 
particularly if the exit probability is low.)  
Then $r_2$ is used to pick uniformly one of the 4 nearest-neighbor 
spins of the flipped spin and this spin is flipped.  
Finally, $r_3$ is used to pick uniformly one of the 
$N$$-$$8$ spins, not including the two flipped spins or their 
6 nearest-neighbor spins, and this spin is flipped.  
This gives the exiting spin configuration for this 
particular absorbing state.  
A similar procedure, using as many random numbers as required, 
is performed if a different state is the absorbing state.  

The time $m$ to exit the absorbing Markov chain is given by 
using 
a random number ${\bar r}$ 
and 
Eq.~(\ref{EAMC3}) 
with $m$ satisfying 
\begin{equation}
\label{EIMCAMCt}
{\vec v}_{\rm I}^{\rm T} {\bf T}^{m} {\vec e} 
< {\bar r} \le 
{\vec v}_{\rm I}^{\rm T} {\bf T}^{m-1} {\vec e} 
\; .  
\end{equation}
Remember that $m$ is in units of Monte Carlo steps (mcs), while 
the physical time is in units of Monte Carlo steps per spin (MCSS).  
This equation does not simplify to an equation 
in closed form as it did for the $s$$=$$1$ case, 
Eq.~(\ref{Einfoldm}).  However, it can be solved iteratively 
using for example a bracketing and bisection method \cite{PRES86}.  
It is critical for issues of algorithmic speed to 
obtain a reasonable guess for $m$ and brackets for $m$ for a 
given absorbing Markov chain and a given ${\bar r}$.  
This can be done by assuming a form such as 
Eq.~(\ref{Einfoldm}) with upper and lower bounds 
\cite{MARC64} 
for the 
eigenvalues of the transient matrix.  

The choice of which state the MCAMC algorithm absorbs 
to is given 
using a random number ${\tilde r}$ and 
the vector ${\vec q}$ of Eq.~(\ref{EAMC5}).  
If the elements of ${\vec q}$ are given by $q_i$, then the 
absorbing state $j$ is the solution of the equation 
\begin{equation}
\label{EIMCAMCR}
\sum_{i=1}^{j-1} q_i \le {\tilde r} < 
\sum_{i=1}^{j} q_i 
\end{equation}
with the first sum being set to zero if $j$$=$$1$.  
This corresponds to Eq.~(\ref{EnfoldR}) for the 
$n$-fold way algorithm.  
For the particular 
$s$$=$$2$ MCAMC of 
Eqs.~(\ref{Es2IT}) and (\ref{Es2IR}), 
the order of the calculation
of $m$ and the exit state is not important, just as the 
order is not important for the $n$-fold way algorithm.  
This is because there is only one transient state that the 
absorbing Markov chain can exit from.  In this case 
the explicit $m$ dependence in Eq.~(\ref{EAMC5}) cancels 
in the numerator and denominator.  
However, in general this is not the case and 
Eq.~(\ref{EAMC5}) 
and Eq.~(\ref{EIMCAMCR}) 
must be used to calculate the absorbing state one must exit to.  

If a MCAMC algorithm for fixed $s$ is not performing 
very well, it can always be improved upon by adding more 
states to the transient 
subspace.  A natural question is which state(s) should be added 
first.  The more transient states there are, the longer 
calculations using Eq.~(\ref{EAMC3}) 
and Eq.~(\ref{EAMC5}) will take --- 
as well as there being a higher probability that a programming 
error is made in constructing either the Markov matrix or the 
exiting spin configuration.  
So the states should be added to the transient subspace in 
an intelligent fashion.  The prescription for finding which 
absorbing state(s) should be added to the transient subspace 
is found using Eq.~(\ref{EAMC7}).  Whichever state(s) have the 
largest probabilities of absorption should be added to 
the transient subspace.  Of course this will depend on the 
model and the model parameters --- for example in metastable 
decay for the Ising model the absorbing state with the 
highest probability in Eq.~(\ref{EAMC7}) depends on 
$L$, $|H|$, and $T$.  

For very low temperatures, 
the value of $m$ for each exit can be very large.  
It is then important not to let the discreteness of the random number 
used to obtain $m$ affect the final answer for $m$.  This is important 
for both low-temperature standard Monte Carlo in which the $p_i$ can be 
very small, as well as for all $s$-state MCAMC algorithms.  In practice, 
I accomplished this by using another random number distributed uniformly 
between $0$ and $0.01$ whenever the first chosen random number was 
less than $0.01$.  The factor of $0.01$ is arbitrary and will depend on 
the number of bits of the random number generator in use.  
Furthermore, to go to very low temperatures (and hence large 
lifetimes) I found it useful to code using routines that can 
be set to a large precision.  I used the package MPFUN \cite{MULTPREC}.  

Fits to the required CPU times can be obtained at low temperatures 
by using the energy difference between the nucleating droplet and the 
most probable absorbing state.  For the data in 
Fig.~\ref{figHItime1}, these are listed in Table~2.  

\begin{table}
\label{TableII} 
\begin{tabular}{|c||c|c|r|l|} 
\hline
\hline
Algorithm &  $\Gamma_0$ & $\Gamma$$-$$\Gamma_0$ &  $A$ &  
$J$$/$$k_{\rm B} T$ of fit point              \\
\hline
\hline
 Monte Carlo & 0   & 24$J$-14$|H|$ &  1.21$\times$$10^{-8}$ & 1.231 \\
\hline
$n$-fold way ($s$$=$$1$) & 8$J$-2$|H|$ & 16$J$-12$|H|$ &  
1.32$\times$$10^{-7}$ & 2.1 \\
\hline
 $s$$=$$2$ MCAMC &  12$J$-4$|H|$ & 12$J$-10$|H|$ &  
1.44$\times$$10^{-6}$ & 2.857 \\
\hline
\hline
\end{tabular}
\caption{
The values used to fit the required CPU time for different 
MCAMC algorithms in Fig.~\protect\ref{figHItime1}.  
The fit functions are valid for $\ell_c$$=$$3$, 
so ${2\over 3}J$$<$$|H|$$<$$J$.
}
\end{table}

Average lifetimes using $s$$=$$2$ and $s$$=$$3$ MCAMC 
are presented in 
Figs.~\ref{figIsingTau1}(a) and \ref{figIsingTau2}.  
The CPU time required for 
these MCAMC algorithms for $|H|$$=$$0.75J$ is shown in Fig.~\ref{figHItime1}.  
For $|H|$$<$$2J$ and low temperatures, the higher $s$ MCAMC algorithms can 
perform many orders of magnitude faster than the $s$$=$$1$ MCAMC 
algorithm, which is itself many orders of magnitude faster than 
the standard dynamic Monte Carlo algorithm!  
For example, at $k_{\rm B}T$$=$$0.25$$J$ the $s$$=$$3$ MCAMC algorithm 
is about $10^{16}$ times faster than the standard algorithm!  

\subsection*{5.4~Projective Dynamics for the Ising Model}
\label{Sec-IsingPD}

The MCAMC algorithm performs extremely well in the 
low-temperature and strong-field regime, i.e.\ in the 
single-droplet (SD) and coexistence regimes at 
low temperatures.  However, MCAMC with small $s$ does not perform very 
well at higher temperatures or for large system sizes.  
This is because the average time increment $m$ given by 
Eq.~(\ref{EIMCAMCt}) is 
relatively small for large $T$ or large $L$.  
In principle it would be possible to construct higher $s$ MCAMC 
that would work in this regime, but this would require a 
lot of bookkeeping of the large number of transient and 
absorbing states.  Is there another advanced simulation method that 
can be used in the multi-droplet regime?  
For the simple ion-channel model of Sec.~4.4, 
the solution was found by using the growing and shrinking probabilities.  
However, for an $N$-spin Ising model the underlying Markov matrix 
is $2^N$$\times$$2^N$ with complicated transition probabilities 
between the $2^N$ states.  
In continuous time the probability $P(\{\sigma\},t)$ of the 
$N$ Ising spins being in state $\{\sigma\}$ at time $t$ is given by 
\begin{equation}
\label{EIPD1}
{{\partial P(\{\sigma\},t)}\over{\partial t}} =
\sum_{ \{ \sigma'\} } \Gamma(\{\sigma\}|\{\sigma'\}) P(\{\sigma'\},t) 
\; ,
\end{equation}
where 
$\Gamma(\{\sigma\}|\{\sigma'\})$ is the probability per unit time of 
moving to spin configuration 
$\{\sigma\}$, given that the current spin configuration 
is $\{\sigma'\}$.  

\begin{figure}[t]
\includegraphics[width=.60\textwidth]{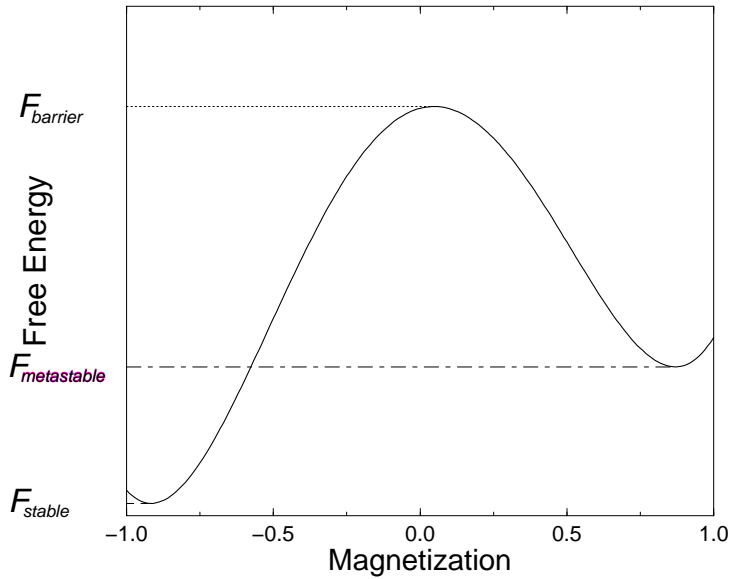}
\caption[]{
A schematic of the free energy per spin 
as a function of the total magnetization $M$ for an Ising model.  
Shown are the metastable well, the stable well, and the 
barrier between the two wells.  
}
\label{figIPD0}
\end{figure}

The idea behind the Projective Dynamics (PD) method is that 
one still expects that a 
simple one-dimensional physical picture 
of the free energy, 
Fig.~\ref{figIPD0}, might hold in this complicated situation.  
In particular, one considers lumping together all states with a 
total magnetization $M$ to form a $(N$$+$$1)$$\times$$(N$$+$$1)$ 
lumped Markov matrix.  
Mathematically, the Markov matrix with $2^N$ states is not 
even weakly lumpable \cite{ABDE82}, but is most likely 
almost weakly lumpable with respect to the 
states along the escape from metastability.  
Thus for the Master equation on the $N$$+$$1$ states 
of a particular magnetization we get 
\begin{equation}
\label{EIPD2}
{{\partial P(M,t)}\over{\partial t}} =
\sum_{ \{ M'\} } W(M|M') P(M',t) 
\; .  
\end{equation}
Note that due to the single-spin flip dynamic there will only 
be non-zero transition probabilities between states that 
have magnetizations that differ by one overturned spin.  
Then the question is how to define 
the lumped transition probabilities 
$W(M|M')$.  One possibility is to assume an 
Arrhenius form \cite{AR1889}
for the transition probabilities for the lumped states 
\cite{LEE95,SCHU80}.  This will give the correct 
free energy for the static model, and actually 
gives a reasonable approximation for $\langle\tau\rangle$ 
in the single-droplet regime \cite{LEE95,SCHU80}.  
It has been extended to projection onto two dimensions, 
magnetization and energy \cite{SHTE97,SHTE99}.  
However, in the multidroplet regime it is 
important to also get the growth phase 
correctly, and this is a non-equilibrium phenomenon.  
This leads to poor agreement between the 
average lifetime obtained from standard simulations and 
by using the Arrhenius form for the 
shrinking and growing probabilities on the 
lumped states \cite{LEE95}.  

However, it is possible to 
actually measure the lumped transition probabilities 
during a sequence of $K$ simulated escapes from 
the initial all-spin-up configuration.  
The lumped matrix is tridiagonal due to the 
single-spin-flip character of the dynamic.  
Define $\langle c_i\rangle_M$ to be 
the average number of spins in class $i$ 
during the simulated escape, 
given that the configuration has magnetization $M$.  
In terms of the flipping probability of a spin in class $i$, 
the growing 
rate of the stable phase (spins down) is 
\begin{equation}
\label{EIPDg}
g(M) = \sum_{i=1}^5 \langle c_i\rangle_M p_i 
\; ,
\end{equation}
and the shrinking rate of the stable phase is 
\begin{equation}
\label{EIPDs}
s(M) = \sum_{i=6}^{10} \langle c_i\rangle_M p_i
\; .  
\end{equation}
Note that the $\langle c_i\rangle_M$ should be measured 
during the actual escape from the initial spin configuration.  
Then the `average' transition probability for the 
lumped matrix is
\begin{equation}
W(M|M') = s(M') \delta_{M+2,M'} + g(M') \delta_{M-2,M'} 
\; .
\end{equation}
In particular, the probability of going from lumped state 
$M$ to lumped state $M'$ is given by 
\begin{equation}
\label{EIPD3}
W(M|M') = {1\over{n(M')}} 
\>\sum_{ \sigma\in\{\sigma|M\} }
\>\sum_{ \sigma'\in\{\sigma|M'\} }
\>\Gamma(\{\sigma\}|\{\sigma'\})
\; ,
\end{equation}
where $n(M)$ is the number of states with magnetization $M$.  

The random walk starts at $M$$=$$L^2$ and terminates 
at $M$$=$$0$.  
We define $h(M)$ as the total time spent in $M$.  
If $L$ is even, 
$h(M)$$=$$0$ for $M$ odd, since these magnetizations are 
not possible.  
Then, in analogy with Eqs.~(\ref{EIonPD15}) and (\ref{EIonPDhi}), 
one has 
\begin{equation}
\label{EIPD4}
\langle\tau\rangle = \sum_{M=2}^{L^2} h(M), \>\>\>
h(M) = {{1+s(M-2)h(M-2)}\over{g(M)}}, \>\>\> 
h(2)= {{1}\over{g(2)}}
\; .  
\end{equation}

An error estimate for 
$\langle\tau\rangle$ can be obtained from error estimates of the 
$h(M)$.  Assume error estimates for $g(M)$ and $s(M)$ have 
been obtained, for example by using a jack-knife procedure 
\cite{BERG92A}.  Then using standard error propagation \cite{TOPP72}, 
the error in the lifetime is 
\begin{equation}
\label{EIPD4E0}
(\delta\tau)^2 = 
\sum_{M=2}^{L^2} \left[\delta h(M)\right]^2
\end{equation}
with 
\begin{equation}
\label{EIPD4E1}
\delta h(2)= {{\delta g(2)}\over{g(2)^2}}
\end{equation}
and iteratively 
\begin{eqnarray}
\label{EIPD4E2}
{\delta h(M)} & = & 
{1\over{g(M)^2}} 
\left\{ 
\left[ 1-s(M-2)h(M-2)\right ]^2\left(\delta g(M)\right)^2
\right .
\nonumber \\
& & +
\left[h(M-2)\right]^2\left(\delta s(M-2)\right)^2
\nonumber \\
& & +
\left . 
\left[s(M-2)\right]^2\left(\delta h(M-2)\right)^2
\right\}^{1\over 2}
\; .  
\end{eqnarray}

How close to the actual lifetime is the average lifetime 
measured by using 
Eq.~(\ref{EIPD4})?  The answer is 
that they are {\it identical\/} to within statistical 
errors!  This is true even though the original Markov matrix 
on $2^N$ states is not lumpable.  
For the Ising model this can be proven using 
class populations 
\cite{NOVO99}.  
However, there is a more transparent proof \cite{MITCUN}.  
Assume that all the $g(M)$ and $s(M)$ were correct.  
Since $g(2)$ is correct, from Eq.~(\ref{EIPD4}) 
$h(2)$ is correct.  
Since $h(2)$, $s(2)$ and $g(4)$ are correct, from 
Eq.~(\ref{EIPD4}) 
$h(4)$ is also correct.  
This can then be iterated to obtain the exact value of 
$\langle\tau\rangle$.  
Note that although the exact value of $\langle\tau\rangle$ 
is obtained, only approximations to the higher moments of 
$\tau$ are obtained.  

\begin{figure}[t]
\includegraphics[width=.30\textwidth]{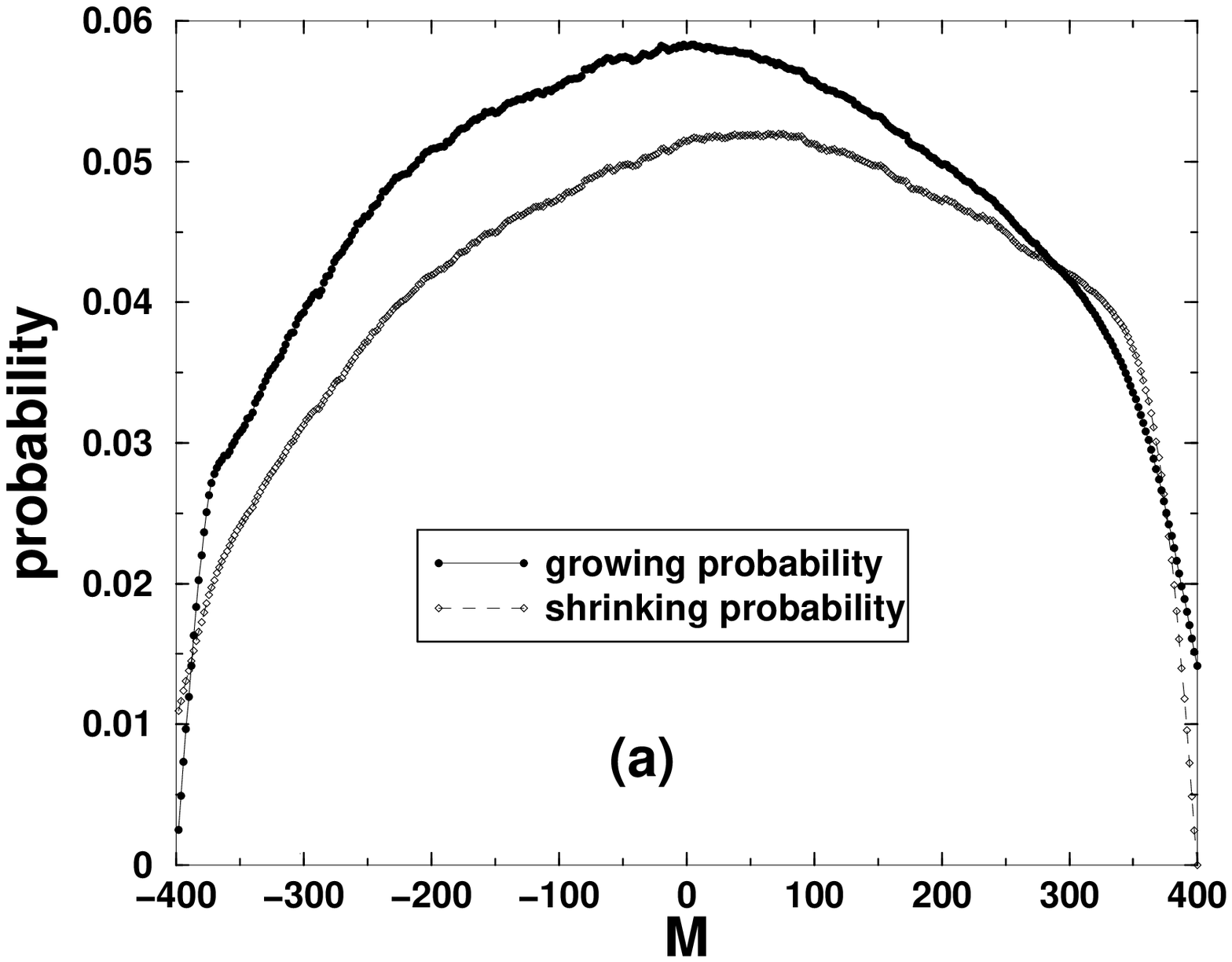}
\includegraphics[width=.30\textwidth]{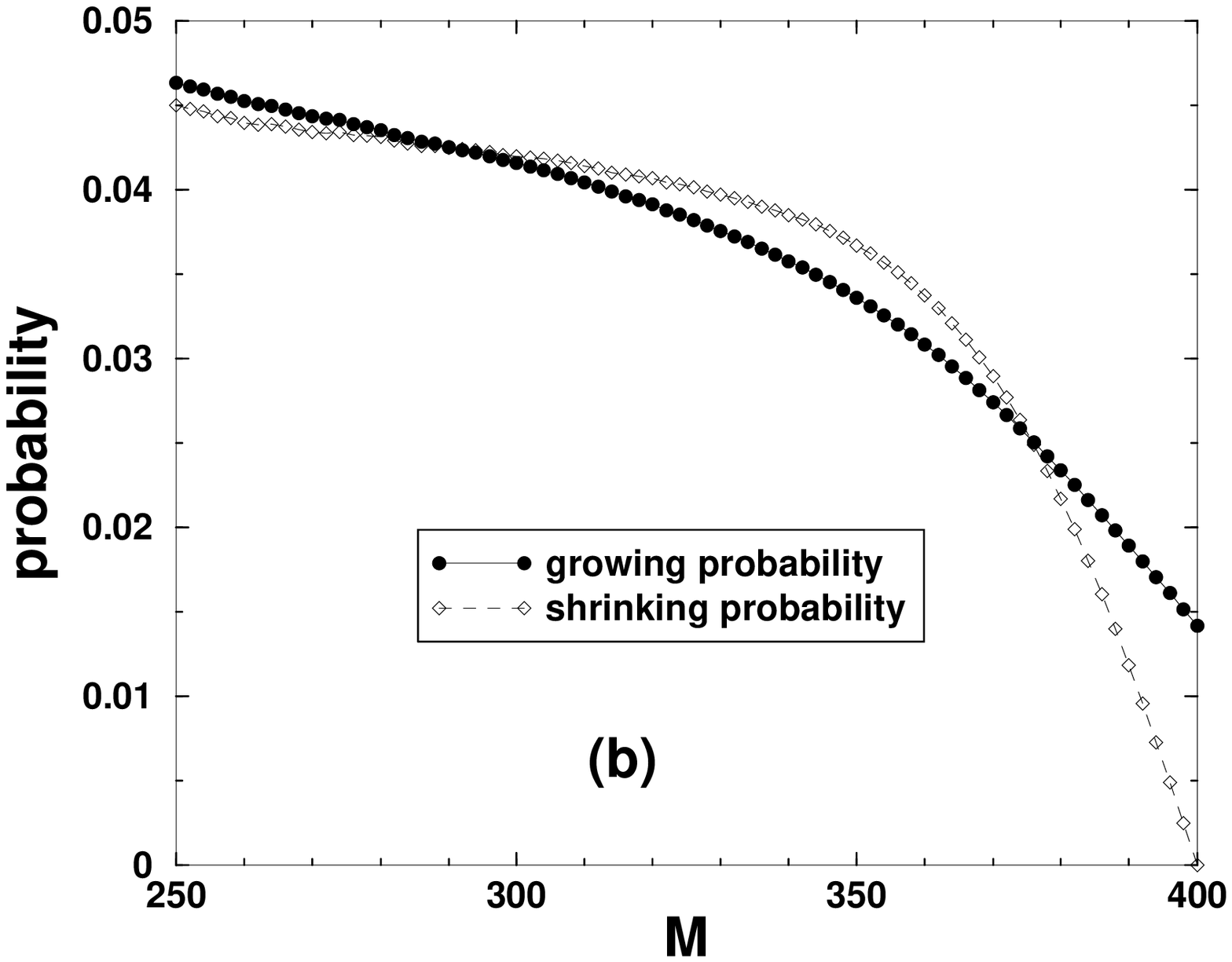}
\includegraphics[width=.30\textwidth]{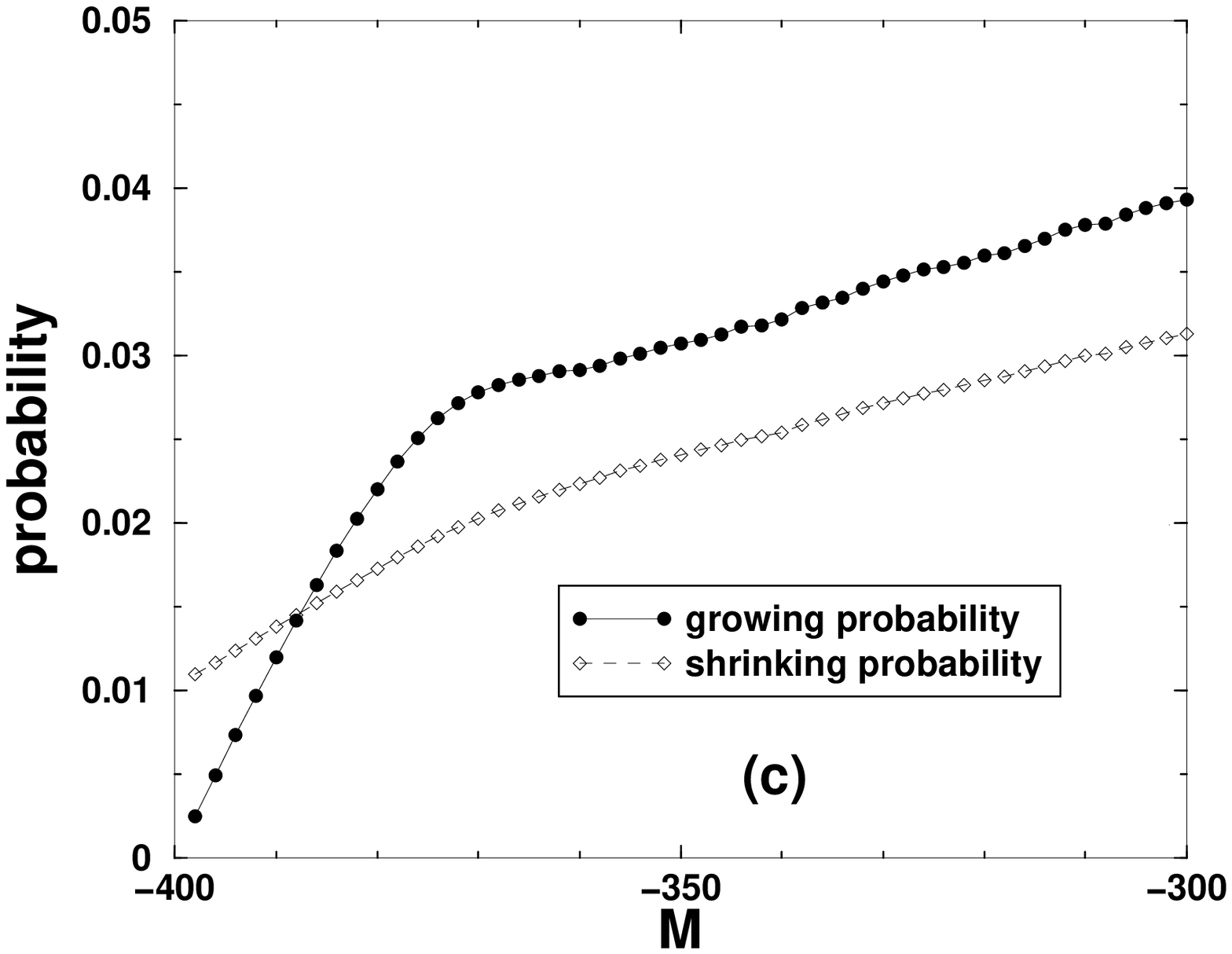}
\caption[]{
The growing and shrinking probabilities for escape of the 
square-lattice Ising model are shown as functions of 
the total magnetization.  Wherever the shrinking and growing 
probabilities are equal, the system is at an extremum of the 
free energy.  These correspond from right to left in 
(a) to the 
metastable phase, the saddle point, and the stable phase.  
(b) Shows the portion near the metastable phase and 
saddle point.  
(c) Shows the portion near the equilibrium phase.  
The parameters are $L$$=$$20$, $T$$=$$0.8$$T_{\rm c}$$\approx$$1.815$$J$, 
$H$$=$$-0.15$$J$.  
The simulation is started with $M$$=$$L^2$ and is run until the 
first passage to $M$$=$$-L^2$.  
}
\label{figIPDgs}
\end{figure}

\begin{figure}[t]
\includegraphics[width=.60\textwidth]{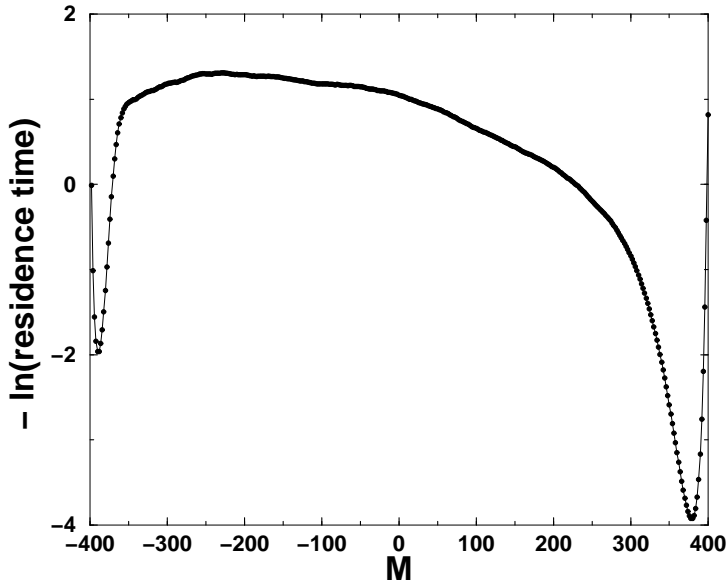}
\caption[]{
The negative of the natural logarithm of the 
residence time, $h(M)$, as a function of 
the system magnetization $M$.  
The parameters are the same as in Fig.~\protect\ref{figIPDgs}.  
The integral of the residence times gives the average lifetime, 
which for these parameters is $1.1$$\times$$10^3$~MCSS.  
This should be compared with 
Fig.~\protect\ref{figIPD0}.  
Note that the stable phase is less probable here than in 
Fig.~\protect\ref{figIPD0} since the simulation is stopped the 
first time the magnetization reaches $M$$=$$-L^2$.  
}
\label{figIPDh}
\end{figure}

The lumped growing and shrinking probabilities also allow one to 
numerically investigate the `free-energy' landscape during the 
escape from the metastable state 
\cite{KOLE98A,KOLE98C}.  
Figure~\ref{figIPDgs} shows these growing and shrinking 
probabilities for a chosen set of parameters for the 
square-lattice Ising model.  
Wherever $g(M)$$=$$s(M)$, 
the probability of growing is equal to the probability of 
shrinking, and that value of $M$ must correspond to 
an extremum during the motion.  In terms of the 
coarse-grained `free energy' during this escape, these correspond to the 
metastable state, the saddle point, and the stable state.  
In more complicated models 
the locations where $g(M)$$=$$s(M)$ also represent extrema.  
This fact allowed 
\cite{KOLE98C} 
both the Barkhausen volumes and 
activation volumes during thermally assisted 
domain-wall motion in a model for 
magnetization reversal in Fe 
sesquilayers on W(110) to be obtained \cite{KOLE97}.  
Using Eq.~(\ref{EIPD4}) to obtain $h(M)$ 
and assuming that just as in equilibrium the `free energy' during 
the escape from the metastable state is 
\begin{equation} 
{\cal F}(M) \propto 
- \ln\left[h(M)\right] 
\end{equation}
gives a way to measure the 
`constrained free energy during escape from the metastable state,' 
Fig.~\ref{figIPDh}, 
similar to 
Fig.~\ref{figIPD0}.  

\begin{figure}[t]
\includegraphics[width=.60\textwidth]{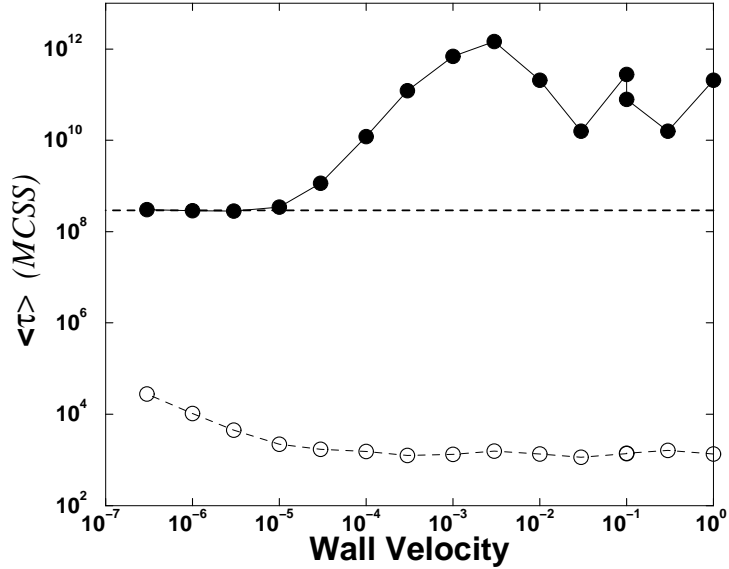}
\caption[]{
The average lifetime $\langle\tau\rangle$ 
measured from the growing and shrinking probabilities 
as a function of the forcing speed is 
shown by the filled symbols for 
a hard forcing wall ($\bullet$).  
The corresponding open symbols ($\circ$) 
are the measured number of Monte 
Carlo steps per spin (MCSS) to escape from the metastable well 
with the forcing present.  The lines connect the data points.  
The heavy dashed line corresponds to the result for 
$\langle\tau\rangle$ from standard simulations.  
These results are for $L$$=$$20$, 
$k_{\rm B} T$$=$$0.5$$J$, and $|H|$$=$$0.75$$J$.  
Note that 
once $\langle\tau\rangle$ is approximately independent of 
the forcing speed, 
there is a difference of about $10^5$ between 
the results measured by the growing and shrinking 
probabilities and the actual escape time with the 
forcing wall.  This $10^5$ difference reflects the 
efficiency of the projective dynamics with a forcing 
wall for these parameters.  
}
\label{figIPD4}
\end{figure}

It is also possible to introduce a moving wall in the magnetization 
to force the system out of the metastable state, and to measure 
$g(M)$ and $s(M)$ during this (hopefully quasi-static) process 
\cite{KOLE98B}.  
The wall position is given by 
\begin{equation}
M_{\rm wall}(t) = (L^2\!+\!1) - v_{\rm wall}  t
\; .  
\end{equation} 
For a soft wall the normal Monte Carlo flip probability is 
multiplied by 
\begin{equation}
p_{\rm wall} = \exp\left[-c\left(M_{\rm new}-M_{\rm wall}\right)\right]
\end{equation} 
if the wall at position $M_{\rm wall}$ is past the 
magnetization of the Monte Carlo new trial configuration 
\cite{NOVO00}.  
Otherwise there is no change in the flip probability.  
Another possibility is to introduce a hard wall, $c$$=$$\infty$, that 
always flips a chosen up spin if $M_{\rm wall}$ is past the current 
magnetization.  
Fig.~\ref{figIPD4}
shows an example of 
lifetimes for a hard forcing wall 
as a function of the wall speed.  
The difference between the lifetimes 
measured from Eq.~(\ref{EIPD4})
with the number of Monte Carlo Steps per spin before escape with the wall 
gives an estimate for the speed-up due to incorporating the 
forcing wall.  Of course, for a wall that moves too fast the process is 
not quasi-static, and the lifetimes are not accurate.  However, 
good results are obtained even for relatively fast walls.  
For example, 
for the hard wall of Fig.~\ref{figIPD4} 
a velocity of $3$$\times$$10^{-6}$M/$L^2$MCSS gives 
a lifetime comparable to an actual escape, but requires almost 
five orders of magnitude less computer time.  
The soft walls do not seem to help the convergence, and 
additional computations are needed at each step to calculate $p_{\rm wall}$.  
This often makes the hard wall more 
computationally efficient than the soft walls \cite{NOVO00}.  
For example, for the same parameters as in 
Fig.~\ref{figIPD4} with $c$$=$$0.01$ 
with a speed of 
$1.0$$\times$$10^{-5}$M/$L^2$MCSS 
the lifetime was $2.6$$\times$$10^{9}$~MCSS, had not yet become 
independent of the forcing speed, and 
required about 2.1 minutes of Cray SV1 CPU time for each 
escape.  This should be compared with the same forcing speed in 
Fig.~\ref{figIPD4} for the hard wall, which required only 
about $0.06$ Cray SV1 CPU minutes for each escape.  

In some cases the class populations $\langle c_i\rangle$ 
during escape from the metastable state can be approximated by 
the class populations in equilibrium, $\langle c_i\rangle^{\rm equ}$ 
\cite{KOLE98A}.  
These equilibrium class populations can be measured during an 
equilibrium Monte Carlo simulation \cite{LEE95}.  
Then using Eq.~(\ref{EIPD4}) with this assumption 
the lifetime can be obtained 
at {\it all\/} temperatures and fields since the 
dependence of the flip probabilities on $T$ and $H$ are 
known.  The approximation for the $\langle c_i\rangle$ 
actually becomes better as $|H|$ becomes smaller since 
then the nonequilibrium configurations play less of a role 
in the metastable decay.  In this way 
lifetimes of about $10^{30}$ Monte Carlo Steps 
per spin have been obtained 
for Ising ferromagnets in two and three dimensions for small $H$ 
\cite{KOLE98A}.  
In addition, with the assumption that the class populations 
do not change with a change of dynamics, values of 
$\langle c_i\rangle$ obtained using one dynamic [say 
Eq.~(\ref{EIGlauber})] 
can be used to obtain lifetimes using a different dynamic 
[say Eq.~(\ref{EIMetrop})].  

\begin{figure}[t]
\includegraphics[width=.60\textwidth]{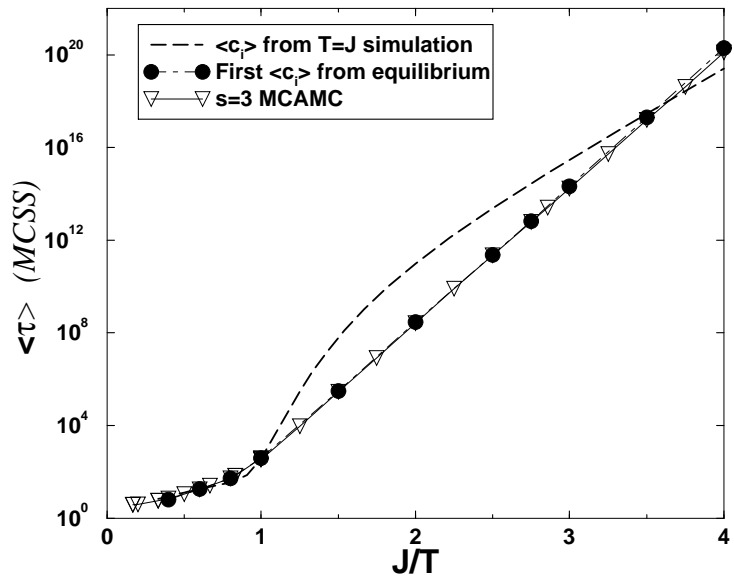}
\caption[]{
The average lifetime, $\langle\tau\rangle$, 
versus inverse temperature 
for $L$$=$$20$ and $H$$=$$-0.75$$J$.  
The $s$$=$$3$ MCAMC results ($\nabla$) are shown for comparison.  
The solid dashed curve is 
from a simulation at $k_{\rm B}T$$=$$J$ 
with the assumption that the spin classes 
are unchanged as $T$ changes.  It has the wrong functional form.  
The bullets ($\bullet$) use this assumption, except the first 
8 overturned spins use class populations 
obtained from equilibrium Monte Carlo simulations at 
fixed magnetization.  
The lines connect the data points.  
}
\label{figIPDtauH}
\end{figure}

As seen in Eqs.~(\ref{EIPDg}) and (\ref{EIPDs}), the growing and 
shrinking rates are functions of both the average of the spin 
classes, $\langle c_i\rangle$, and the single-spin flip 
probabilities, $p_i$.  The functional form for $p_i$ is given by 
Eq.~(\ref{EIGlauber}) using the energies in Table~1.  
Hence the functional form for the $p_i$ are known for all 
$T$ and $H$ values.  
One could assume that the average class 
populations do not change too much with changing field or temperature.  
Then the class populations at one simulated value of $T$ and $H$ can 
be used to obtain growing and shrinking probabilities at other 
values of $H$ and $T$.  
For the parameters in Fig.~\ref{figIPDtauH} this assumption 
produces the wrong functional form for $\langle\tau\rangle$.  
This is because for a growing interface the 
class populations are not constant \cite{RIKV00}.  
An alternative assumption is that the average class populations 
can be given by the simulation results at $k_{\rm B}T$$=$$J$ 
except for magnetizations near the metastable well, where 
class populations from equilibrium Monte Carlo simulations 
\cite{LEE95,KOLE98A} can be used.  
As seen in 
Fig.~\ref{figIPDtauH} this combination of assumptions 
works well for the chosen parameters over almost the 
entire temperature range.  This is because at low 
temperatures the growth phase contributes very little to 
the average lifetime.  Furthermore, for our fixed $L$ and $H$, the 
most probable equilibrium configuration is nearly the 
same as the most probable path of the 
nucleating droplet at fixed $M$ as $T$ becomes small.  
Such techniques must be used in conjunction with knowledge about 
the physics of the escape from the metastable state to be 
accurate, as 
Fig.~\ref{figIPDtauH} demonstrates.  
Simple approximate class populations during the growth phase 
\cite{RIKV00} 
could alternatively be used 
for the growth portion of these types of calculations.  

\begin{figure}[t]
\includegraphics[width=.60\textwidth]{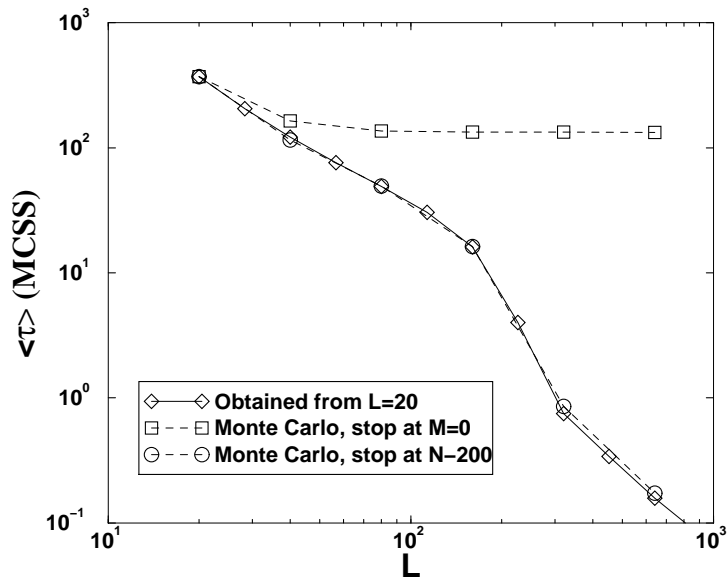}
\caption[]{
The lifetimes obtained from extrapolating in $L$ using 
Eqs.~(\protect\ref{EPDN2N}) and (\protect\ref{EPDN2Ns}) from 
simulations at the smallest lattice size, $L$$=$$20$.  
This is for $H$$=$$-0.75$$J$ and $k_{\rm B}T$$=$$J$.  
When the cutoff for the lifetime is kept constant at 
$M$$=$$0$, once the 
system is in the multi-droplet regime (for large $L$) 
the lifetime is independent of the system size ($\Box$).  
When the cutoff for the lifetime is not kept constant in 
$M$, but is rather constant in the number of overturned 
spins (here at $L^2$$-$$200$) for both the 
Monte Carlo ($\circ$) and extrapolated values ($\Diamond$) 
the lifetime depends on $L$.  
}
\label{figIPDtauV}
\end{figure}

It is also possible to approximate the growing and 
shrinking probabilities of a system of size $2N$ from a system 
of size $N$ \cite{KOLE98A}.  
Since the relevant configurations typically contain many small 
droplets, to a good approximation the larger system can be viewed 
as consisting of two `independent' copies of the smaller system.  
In this approximation one can write that 
\begin{equation}
\label{EPDN2N}
g(2N,M) \approx 
{
{\sum_{M'=M}^{2N} h(N,N+M-M') h(N,M')\left[g(N,N+M-M')+g(N,M'-N)\right]} 
\over 
{\sum_{M'=M}^{2N} h(N,N+M-M') h(N,M'-N)}
}
\end{equation}
and
\begin{equation}
\label{EPDN2Ns}
s(2N,M) \approx 
{
{\sum_{M'=M}^{N} h(N,N+M-M') h(N,M')\left[s(N,N+M-M')+s(N,M'-N)\right]} 
\over 
{\sum_{M'=M}^{N} h(N,N+M-M') h(N,M'-N)}
}
\end{equation}
where the explicit dependence on the volume has been included.  
Remember that for $N$ even, the odd magnetizations can be included 
since they have zero residence times.  
Note that the cut-off in this extrapolation does not remain 
at the original value of $M$, but rather is a constant in the 
number of overturned spins on a lattice of any size.  
Using Eqs.~(\ref{EPDN2N}) and (\ref{EPDN2Ns}) 
repeatedly, one can extrapolate to very large lattices.  
While the lattice sizes extrapolated to are nowhere near 
world record sizes \cite{STAU96}, 
much additional information has been 
obtained from the small-lattice simulations.  
This is illustrated in Fig.~\ref{figIPDtauV}.  

\begin{figure}[t]
\includegraphics[width=.30\textwidth]{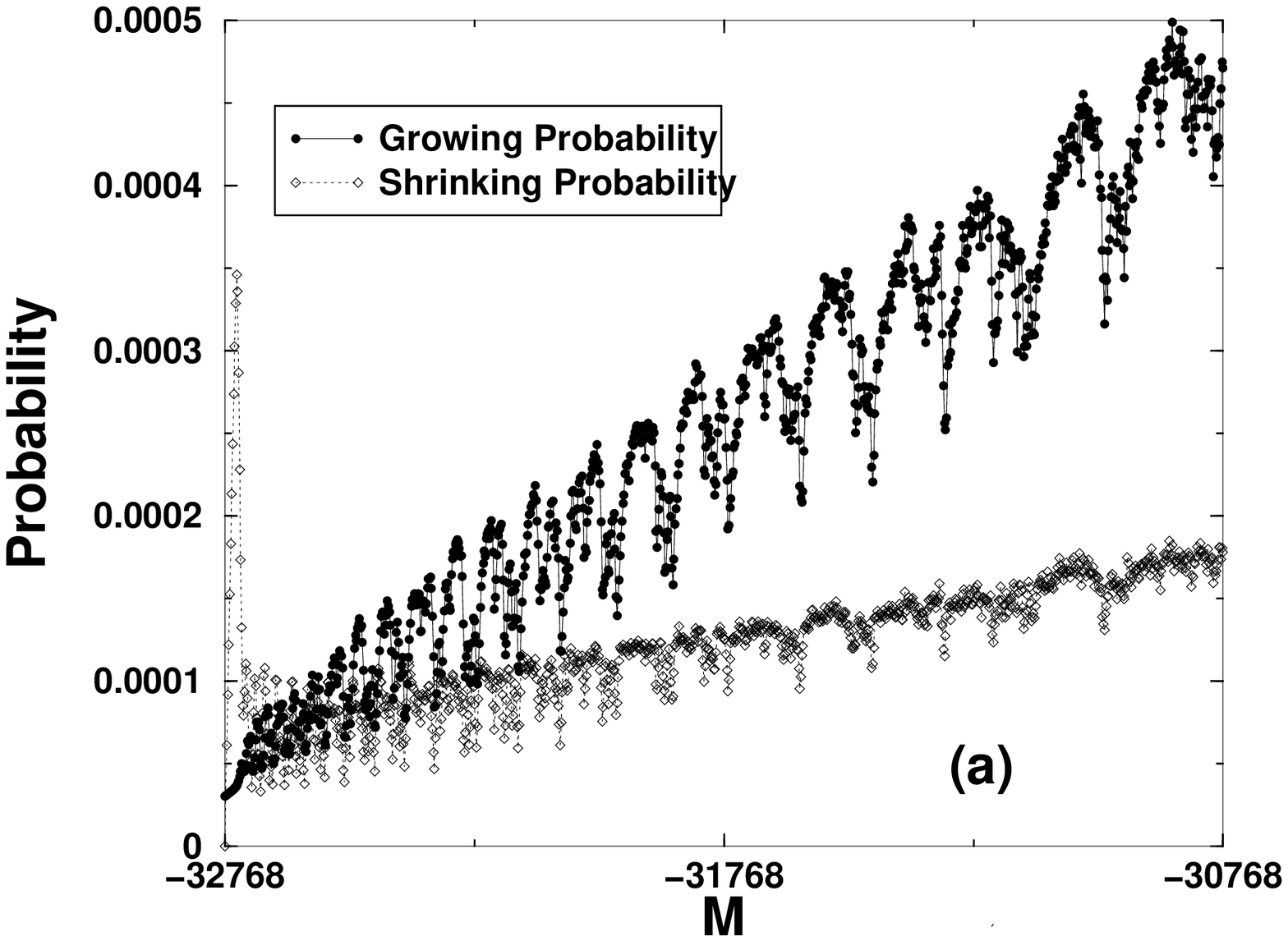}
\includegraphics[width=.30\textwidth]{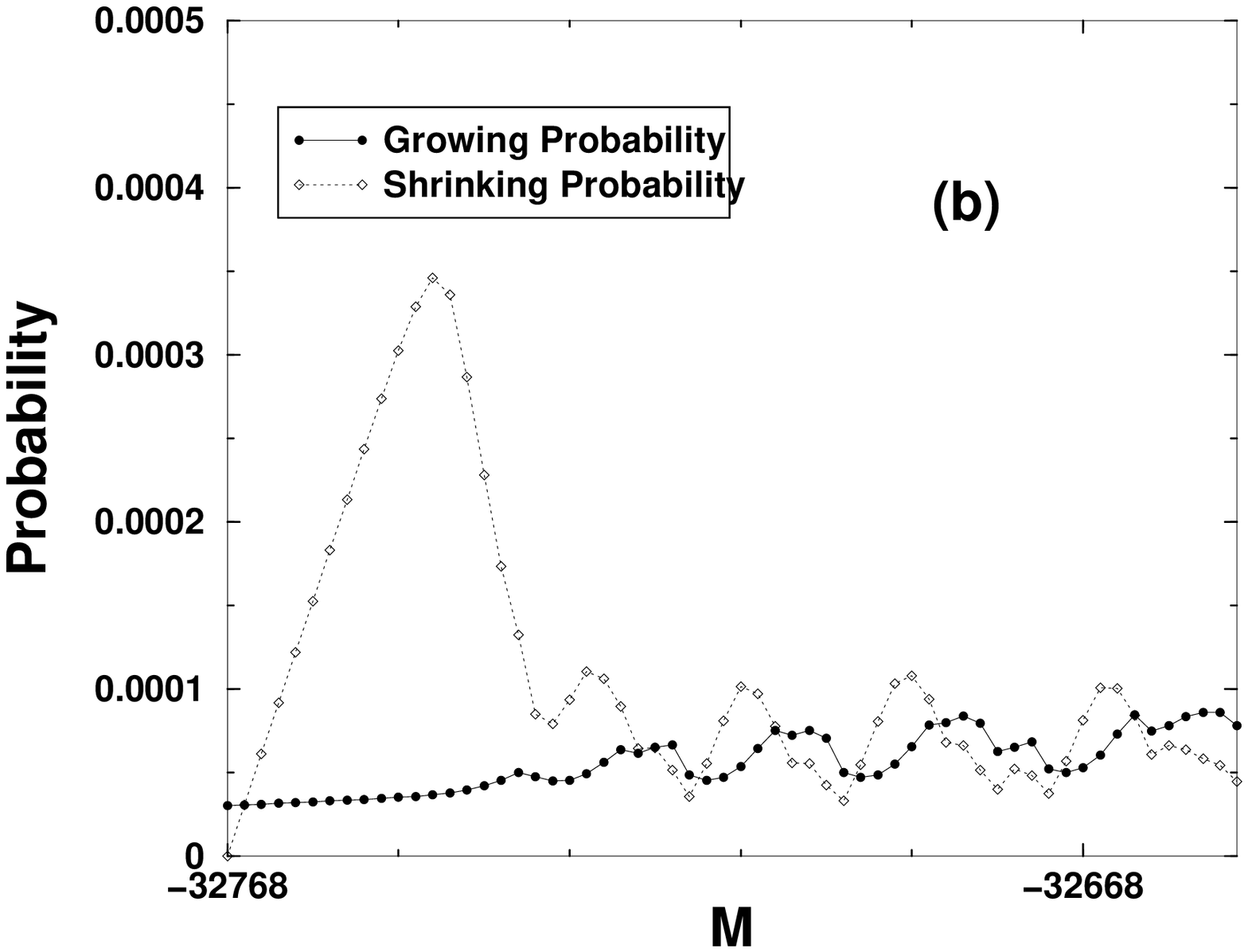}
\includegraphics[width=.30\textwidth]{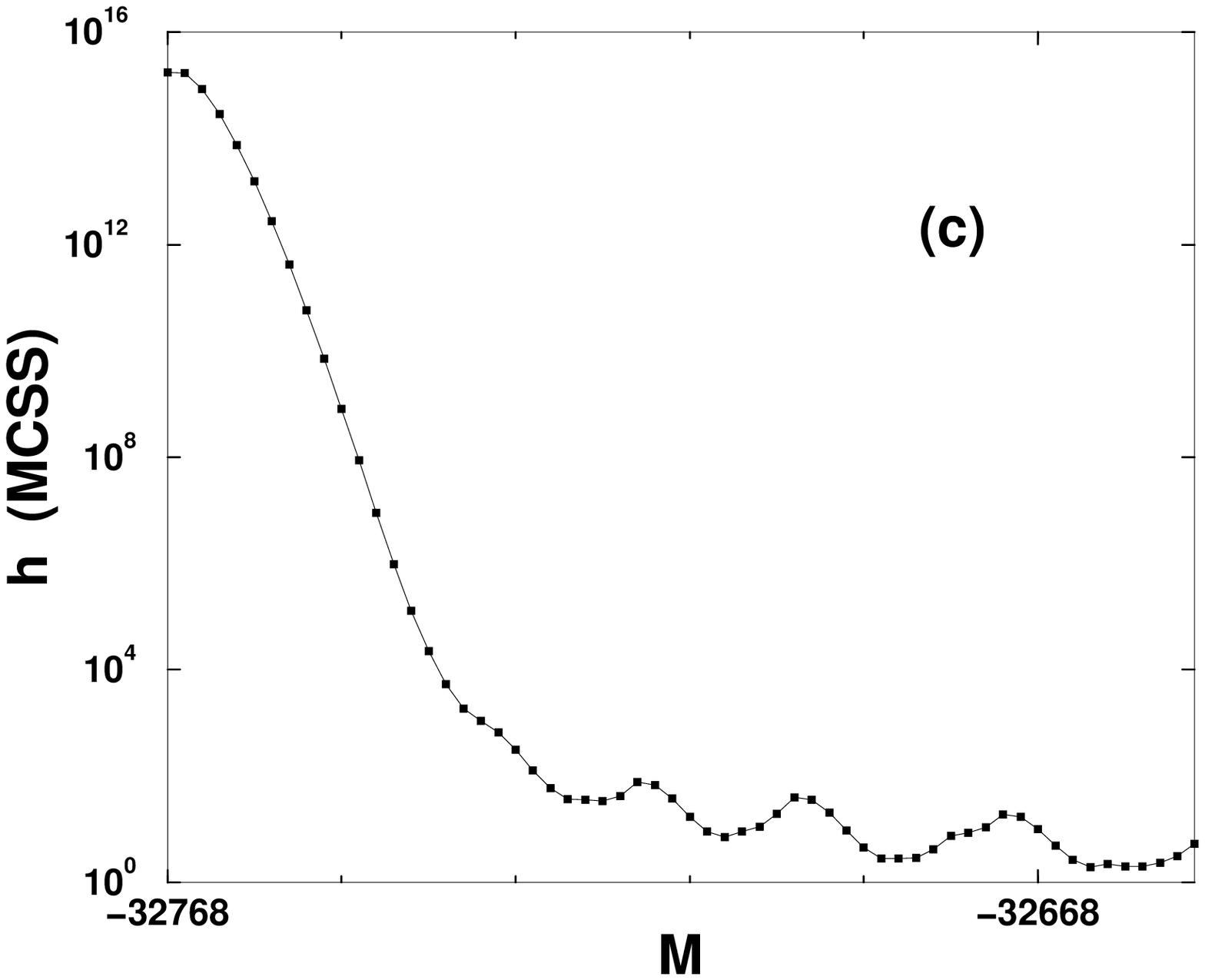}
\caption[]{
Results for a three-dimensional $32^3$ simple cubic Ising model using 
projective dynamics.  This is for a temperature of 
$k_{\rm B}T$$=$$0.9026819$$J$ at $|H|$$=$$1.3$$J$ and using a 
forcing speed of $5.0$$\times$$10^3$ (for units see the text).  
These are plotted as a function of the total system magnetization.  
The average lifetime for this escape is 
$\langle\tau\rangle$$=$$7.3$$\times$$10^{15}$~MCSS.  
(a) The growing and shrinking probabilities up to 
1024 overturned spins.  
(b) A portion of (a) showing the growing and shrinking probabilities 
near the metastable state.  
(c) The residence time obtained using these growing and 
shrinking probabilities.  
}
\label{figIPD3d}
\end{figure}

These types of projective dynamics techniques with a forcing wall 
can also be used on other discrete spin models and on other 
lattices to obtain extremely long lifetimes \cite{KOLEUN}.  
For example, Figure~\ref{figIPD3d} shows results for the 
simple-cubic Ising ferromagnet on a $32^3$ lattice.  The 
temperature and field are such that 
$\langle\tau\rangle$$=$$7.3$$\times$$10^{15}$~MCSS.  
This lifetime is more than $10^6$ times longer than a recent 
large-scale simulation of the model using standard Monte Carlo 
simulations \cite{STAU99}.  
Here the inverse speed of the hard forcing wall was 
$5000$ $n$-fold way updates per spin per overturned spin.  
For the square lattice at low temperatures at fixed $L$, the nucleating 
droplet has neither a square nor a circular shape 
\cite{LEE95,GUNT93,GUNT94}.  
Similarly, for the simple cubic lattice 
the nucleating droplet at low temperatures 
for fixed $L$ has a particular shape which is neither a cube 
nor a sphere \cite{CHEN97A,CHEN97B}.  These discrete lattice effects 
cause the oscillations in the quantities seen in 
Fig.~\ref{figIPD3d}.  

\subsection*{5.5~Parallelization of Dynamic Monte Carlo}
\label{Sec-IsingParallel}

With the widespread availability of massively parallel computers, 
with the number of processing elements (PEs) on the order of hundreds to 
thousands, the question arises as to how dynamic Monte Carlo 
simulations can be implemented efficiently on these 
parallel computers.  Often the parallelization can be accomplished 
in a trivial fashion, just by letting each PE run its own copy of 
one of the $K$ escapes.  Then no communication between PEs must be 
performed until the escape is finished and statistics are collected.  
However, sometimes the lattice size to be simulated is too large to 
fit onto a single PE, or fitting the simulation on a PE will use too 
much swapping of virtual memory, leading to long execution times.  
In such cases a non-trivial parallelization scheme is needed.  

\begin{figure}[t]
\includegraphics[width=.60\textwidth]{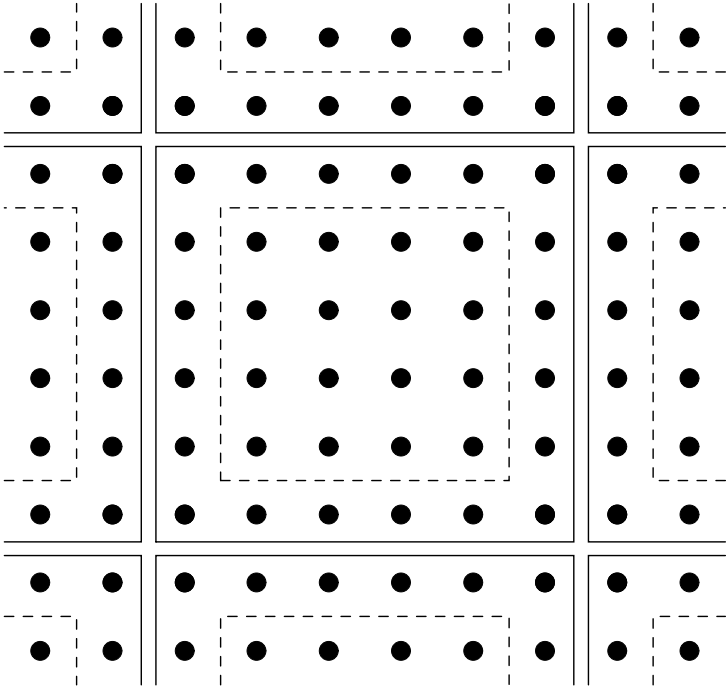}
\caption[]{
A portion of a lattice showing the arrangement of spins 
on a single processing element (PE) and its neighboring PEs.  
This is for $\ell$$=$$6$.  The solid lines 
represent the boundaries of spins on a single processor, while 
the dashed lines isolate the boundary spins from the 
interior spins.  
}
\label{figT3E0}
\end{figure}

The initial impression might be that dynamic 
Monte Carlo is a scalar algorithm.  However, in 1988 
Lubachevsky \cite{LUBA88}
showed that the dynamic Monte Carlo simulation for the Ising 
model belongs to a class of problems called 
discrete-event simulations and can be parallelized.  
As mentioned in Sec.~4.2, 
a discrete-event simulation \cite{LAWK91} is a problem in which the 
state vector changes discontinuously at certain (perhaps random) 
times.  
Discrete-event simulations are used in an extremely wide range 
of applications in science, technology, manufacturing, and sociology.  
Just a few recent examples include 
preventative veterinary medicine \cite{ALLO99}, 
resource allocation following a terrorist bombing \cite{HIRS99}, 
modeling concrete supply and delivery \cite{SMIT99}, 
aircraft design \cite{POWE99}, 
ecological models \cite{NICO77,GLAS97}, 
design of logical circuits \cite{FROH97,KELL99}, 
manufacturing systems \cite{PENG96}, 
battlefield simulations \cite{HILL97}, 
and 
simulation of wireless services \cite{BORS97}.  
For the Ising model 
the method of parallelization is to place a block of spins, 
say $\ell$$\times$$\ell$ in two dimensions, on a processing element.  
Then the total number of spins on the lattice is 
$N$$=$$L^2$$=$$\ell^2 N_{\rm PE}$ where 
$N_{\rm PE}$ is the number of PEs used in the simulation.  
A block of spins and its neighboring spins from nearest-neighbor blocks on a 
square lattice is shown in 
Fig.~\ref{figT3E0}, shown for $\ell$$=$$6$.  
For $\ell$$=$$6$ there are $16$ interior spins and $20$ boundary 
spins on each PE.  

To parallelize the dynamic Monte Carlo algorithm, each PE has 
its own time clock with its own simulation time.  
This local time is called the `virtual time' of the block of spins 
\cite{JEFF85}.  
Remember that the entire system has a 
single time --- but for parallel implementation each PE has its 
own clock with its own virtual time.  
To calculate some quantity at a particular time, 
for example the spin configuration at time $t$, 
the simulation must evolve until the clocks on all the PEs are 
at time $t$.  
However, if one is interested only in advancing the spin 
configuration over time (as opposed to measuring the spin 
configuration at a particular time) each PE can 
independently advance the spins in its block and update its 
time.  

Of course the blocks of spins on nearest-neighbor PEs are 
not independent.  Consequently the parallelization must ensure 
that causality is not violated as the simulation progresses.  
The conservative approach 
\cite{LUBA88}
to avoid causality violations is to have a PE idle (wait) 
until it is guaranteed that an update will not violate 
causality.  If the time on a PE's clock is behind that of 
all of its nearest-neighbor PEs, then there is no 
possibility that causality is violated, and the PE can 
advance its spin configuration and time.  
Another approach to avoiding causality violations, called the speculative or 
optimistic approach, is to assume that no causality violations occur and 
have every PE always advancing its spin configuration 
\cite{JEFF85,FUJI90}.  
Of course at some 
point a causality violation will occur for a particular PE.  Then 
it must roll back to a previous time and spin configuration before 
causality was violated.  This rollback mechanism can then cause 
causality violations and hence rollbacks on the nearest-neighbor PEs.  
Hence such rollbacks can cascade 
[11--114] 
over the lattice.  

We will only describe parallelization of the dynamic 
Monte Carlo for the Ising model using the 
conservative approach \cite{LUBA88}.  
The same approach can easily be applied to other 
dynamic models, including dynamic classical models with 
continuous degrees of freedom.  
Each PE advances both 
its clock and its $\ell$$\times$$\ell$ block of spins 
{\it if and only if\/} its virtual time is not greater than the virtual 
times of {\it all\/} its nearest-neighbor blocks.  
Otherwise the PE idles until its virtual time is 
not greater than those of its nearest-neighbor PEs.  
In other words, if the virtual time for a PE is not 
a local minimum of the `virtual time' surface then 
it executes a {\tt wait until\/} directive 
until its virtual time is a local minimum.  
The `virtual time' surface is the surface of 
`virtual times' on all the PEs.  
This algorithm is free of deadlocks, because at worst 
the PE with the global 
minimum virtual time can advance its virtual time.  

The checking restriction of virtual times can be relaxed 
for $\ell$$>$$1$ 
so that 
once a PE chooses a boundary spin to try to flip it checks 
the virtual time only of 
the single PE 
(or the two PEs for a corner spin) 
that the chosen boundary spin has its nearest-neighbor spin on.  
If this neighboring PE's virtual time is less than the updating PE's 
virtual time, then the updating PE {\tt waits until\/} its 
virtual time is less than this neighboring PE's virtual time 
before the update attempt is made.  

\begin{figure}[t]
\includegraphics[width=.60\textwidth]{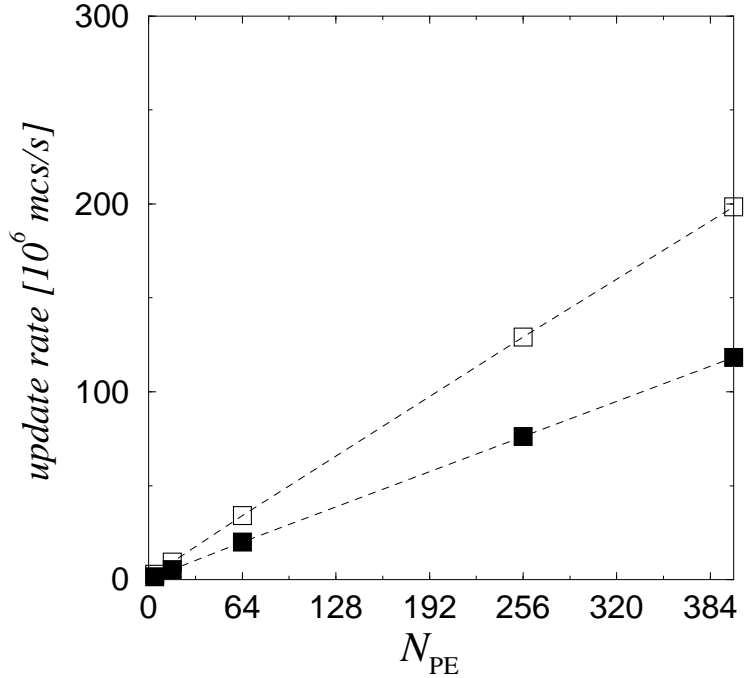}
\caption[]{
The update rate 
for the square-lattice Ising model 
(in units of a million Monte Carlo steps per second of wall-clock time)
is shown as a function of the number of 
processing elements (PEs) of a Cray T3E, $N_{\rm PE}$.  
The algorithm is implemented using a conservative 
discrete-event simulation approach.  
The filled squares are for the standard Monte Carlo 
algorithm, while the open squares are for the shielded 
$n$-fold way algorithm (in continuous time).  
The temperature is $T$$=$$0.7$$T_{\rm c}$ and the applied 
field is $|H|$$=$$0.2857J$.  
Spins in blocks of $\ell$$\times$$\ell$ are placed on each 
PE, with $\ell$$=$$128$.  
The size of the simulated lattice is 
$L$$=$$\ell$$\protect\sqrt{N_{\rm PE}}$, 
so the 
largest lattice simulated has $L$$=$$2560$.  
The lines are guides for the eye.  
Note that a straight line fit would correspond to 
perfect scaling of the parallel algorithm.  
After Ref.~\protect\cite{KORN99A}.  
}
\label{figT3Eup}
\end{figure}

Figure~\ref{figT3Eup} shows the update rate in 
Monte Carlo steps (mcs) per second on a Cray T3E 
as a function of the number of processing elements ($N_{\rm PE}$) 
used.  
This is based on work reported in 
Refs.~\cite{KORN99A,KORN99B}.  
The standard dynamic Monte Carlo algorithm 
of Sec.~5.1 
was run using 
$N_{\rm PE}$ values of $8$, $16$, $64$, $128$, and $400$, 
a fixed block size with $\ell$$=$$128$, 
$T$$=$$0.7$$T_{\rm c}$, and 
$|H|$$=$$0.2857J$.  
The size of the simulated lattice was 
$L$$=$$\ell$$\sqrt{N_{\rm PE}}$, 
so the 
largest lattice simulated had $L$$=$$2560$.  
Figure~\ref{figT3Eup} shows that the algorithm seems to be scalable --- 
in other words, 
the update rate increases linearly with $N_{\rm PE}$ 
as the number of PEs and the lattice size $L$ 
increases for a fixed $T$, $H$, and $\ell$.  

\begin{figure}[t]
\includegraphics[width=.60\textwidth]{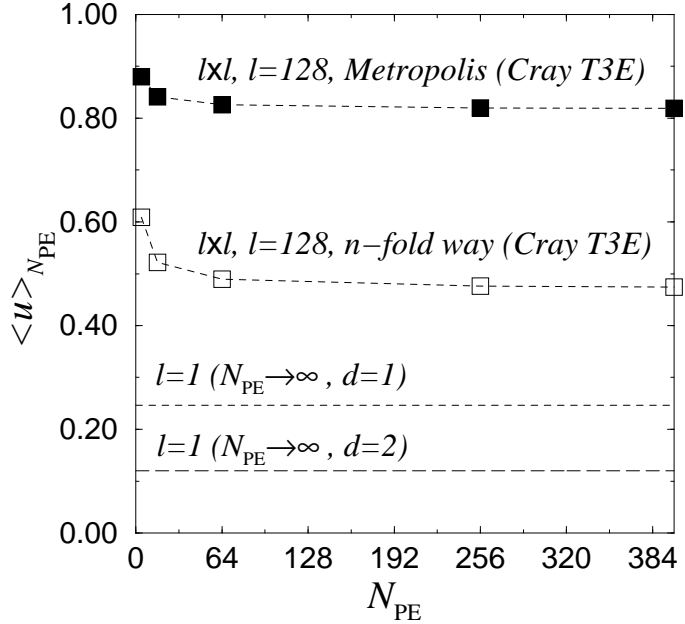}
\caption[]{
The average utilization, $\langle u\rangle$, is shown as a 
function of the number of processing elements (PEs) used.  
The parallelization algorithm uses a 
conservative discrete-event simulation approach 
\protect\cite{LUBA88}.  
The filled squares correspond to the standard dynamic Monte Carlo 
simulation for the square-lattice Ising model 
shown in Fig.~\protect\ref{figT3Eup}, 
where 
the utilization plateaus at about 80\%.  
Here each PE has an $\ell$$\times$$\ell$ block of spins, 
with $\ell$$=$$128$.  
The average utilization for the $n$-fold way algorithm 
plateaus at about 50\%.  
The short-dashed line corresponds to the worst-case utilization 
in the limit $N_{\rm PE}$$\rightarrow$$\infty$ for a 
one-dimensional lattice and $\ell$$=$$1$ (one spin per PE).  
This is obtained from finite-size scaling for various lattice sizes 
\protect\cite{KORN00A}.  
The long-dashed line corresponds to the worst-case utilization 
in the limit $N_{\rm PE}$$\rightarrow$$\infty$ for a 
two-dimensional square lattice 
and $\ell$$=$$1$ 
\protect\cite{KORN00C}, 
with results 
obtained by running large lattice simulations.  
}
\label{figT3Eutil}
\end{figure}

One important question in parallel discrete-event simulations using 
the conservative approach is the utilization of the 
PEs.  In other words, what fraction of the PEs are not idling at 
a particular time?  
This can be answered by using 
non-equilibrium surface science methodologies 
\cite{KORN00A,KORN00B} 
to study the 
average utilization, $\langle u\rangle$, of the parallel algorithm.  
The value of $\langle u\rangle$ for the standard dynamic 
Monte Carlo simulations of Fig.~\ref{figT3Eup} 
are shown in Fig.~\ref{figT3Eutil}.  

For the worst-case scenario for the average utilization, 
$\ell$$=$$1$ (one spin on each block), 
it is possible to study the average utilization using 
finite-size scaling of simulations of the virtual time 
to obtain information about 
$\langle u\rangle$ as $N_{\rm PE}$$\rightarrow$$\infty$ 
\cite{KORN00A,KORN00C}.  
Fig.~\ref{figT3Eutil} shows the result obtained for a 
square lattice \cite{KORN00C} from finite-size scaling 
of simulation data.  
Much more information about the virtual time surface can 
be obtained for a one-dimensional lattice (with periodic 
boundary conditions).  In Ref.~\cite{KORN00A} it was 
shown that the virtual time surface for the conservative 
implementation of parallel discrete-event simulations 
is governed by the Edwards-Wilkinson Hamiltonian 
\cite{EDWA82,BARA95}.  
Universality arguments then prove that 
$\langle u\rangle$ must be bounded away from zero 
as $N_{\rm PE}$$\rightarrow$$\infty$.  This {\it proves\/} 
that in one dimension 
the conservative implementation of parallel 
discrete-event simulations are scalable in the worst-case scenario!  
This is a very general result for all one-dimensional 
lattices, {\it regardless\/} 
of the actual physical problem being simulated.  
In one dimension precise finite-size scaling can be used 
to show that $\langle u\rangle$ is just slightly less than 
${1\over 4}$ in the worst-case scenario.  This value is also 
shown in 
Fig.~\ref{figT3Eutil}.  
The fact that $\langle u\rangle$ is finite, and hence the 
algorithm is scalable, is a non-trivial result.  It has been 
shown, including exact analytic results, 
that there are other surfaces 
from reasonable growth models 
where the density of local minima approaches zero \cite{TORO00}.  
A finite-size scaling analysis of the approach to the 
$N_{\rm PE}$$\rightarrow$$\infty$ limit in one dimension shows that 
\begin{equation}
\langle u\rangle_{N_{\rm PE}} - 
\langle u\rangle_{\infty} \propto {1\over N_{\rm PE}}
\end{equation}
for the $\ell$$=$$1$ case \cite{KORN00A}.  
This is the scaling form for the density of 
minima for this type of nonequilibrium surface \cite{KRUG90}.  
In addition it was shown \cite{KORN00A} 
that the fluctuations in the density of 
local minima is 
\begin{equation}
\sigma_{N_{\rm PE}}^2 
=
\langle u^2\rangle_{N_{\rm PE}} - 
\langle u\rangle_{N_{\rm PE}}^2  
\propto {1\over {\sqrt{N_{\rm PE}}}}
\end{equation}
for the one dimensional $\ell$$=$$1$ case \cite{KORN00A}.  

Another intrinsic property of the virtual time surface 
is its width as $N_{\rm PE}$$\rightarrow$$\infty$.  
This is important because, in order to measure a spin configuration at a 
particular time $t$, all PEs must wait (idle) once they reach 
time $t$ until {\it all\/} the PEs have caught up.  
If the virtual time surface has a width which diverges 
as $N_{\rm PE}$$\rightarrow$$\infty$, then this measurement 
process will not scale.  In Ref.~\cite{KORN00A} it was 
shown that indeed the interface width diverges in 
one dimension since the virtual time horizon belongs to the 
Edwards-Wilkinson universality class.  
In higher dimensions numerical studies also indicate a 
divergence of the width of the virtual time surface \cite{KORN00C}.  
It has, however, been shown that a slight change in the algorithm 
\cite{KORN00B} can produce both a scalable algorithm and 
a virtual time surface with a finite surface width.  

Another way to handle measurements is with a buffer of spin 
configurations.  These spin configurations 
are stored once the virtual time on a PE 
is equal to a particular time at which a measurement is to be made.  
As long as the buffer is finite and the virtual time surface 
width diverges 
as $N_{\rm PE}$$\rightarrow$$\infty$, 
this measurement procedure will also not scale.  

The inner loop of the parallel dynamic Monte Carlo algorithm 
is executed on each PE in parallel.  
Each PE updates an $\ell$$\times$$\ell$ block of spins, 
and each PE has its own clock to keep its virtual time.  
The parallelization is much easier in continuous time since in this 
case no synchronization is required because the virtual times on 
neighboring PEs cannot be equal.  
The simulation is started with all spins up and each PE 
has its virtual time set to zero.  
Then each PE determines the time 
$\Delta t$$=$$-\ln({\bar r})$, where ${\bar r}$ is a 
uniformly distributed random number, of its first update attempt.  
The inner loop (written here specifically for a square lattice) 
is then executed repeatedly:
\begin{itemize}
\item Select a spin from the $\ell^2$ spins on the block with equal 
probability.  
\item If the chosen spin is:
\begin{itemize}
\item[$\circ$] One of the $(\ell$$-$$2)^2$ spins in the kernel of the 
block proceed to the next step.  
\item[$\circ$] One of the $4\ell$$-$$4$ spins on the boundary, 
then check the virtual time of 
the neighboring PE or PEs that the chosen spin interacts with.  
If at least one of the neighboring PEs has a virtual time less than 
the updating PE's virtual time of the next update, then 
{\tt wait until\/} this is no longer the case.  
(For the square lattice, this checking 
and wait-until condition requires 2 PEs for corner spins and 
one PE otherwise.)  
\end{itemize}
\item Update the spin using the same probabilities as in the 
standard dynamic Monte Carlo serial algorithm, i.e.\ with 
Eq.~(\ref{EIGlauber}) or Eq.~(\ref{EIMetrop}).  
\item Add $\Delta t$ to the local virtual time.  
\item Determine the virtual time of the new next update 
by using the {\it local\/} time increment 
$\Delta t$$=$$-\ln({\bar r})$, where ${\bar r}$ is a 
uniformly distributed random number.  
\end{itemize}
The performance of this algorithm is shown in 
Fig.~\ref{figT3Eup}, and its utilization is shown in 
Fig.~\ref{figT3Eutil}.  

Can the advanced dynamic Monte Carlo algorithms also be 
parallelized?  
Lubachevsky in 1988 
\cite{LUBA88} proposed a method to parallelize in a 
conservative fashion the $n$-fold way algorithm.  
This was implemented and tested for the first time in 
1999 \cite{KORN99A,KORN99B}.  
The idea is to have each PE using the $n$-fold algorithm 
with equations similar to 
Eq.~(\ref{Einfoldm}) in discrete time or Eq.~(\ref{EnfoldDt}) 
in continuous time to update its virtual time.  
However, to ensure causality is not violated a 
`shield' of spins at each surface acts as an absorbing state.  
In each block an additional absorbing class is defined which 
contains the spins on the boundary.  There are 
$N_{\rm b}$$=$$4(\ell$$-$$1$$)$ of these spins.  The original 
$n$-fold way tabulation of the $10$ spin classes for the square lattice 
are only performed for the $N_{\rm k}$$=$$(\ell$$-$$2$$)^2$ 
spins in the kernel.  
Hence $N_{\rm k}$$=$$\sum_{i=1}^{10}n_i$ and 
$N_{\rm b}$$+$$N_{\rm k}$$=$$\ell^2$.  
Note that each PE keeps its own class populations and its own virtual time. 
For each escape the simulation 
is started with all spins up and each PE 
has its virtual time set to zero.  
Then each PE determines the time 
$\Delta t$ from Eq.~(\ref{EdtP}) of its first spin update.  
The inner loop of the 
continuous time algorithm, performed by every PE in parallel, is: 
\begin{itemize}
\item Use a uniformly distributed random number 
${\tilde r}$ to select a class according to 
the relative weights 
$\{ N_{\rm b}, n_1 p_1, n_2 p_2, \cdots, n_{10} p_{10}\}$, then use 
another random number $r_1$ to select a spin from the chosen class 
with equal probabilities.   
\item If the spin is:
\begin{itemize}
\item[$\circ$] One of the $(\ell$$-$$2)^2$ spins in the kernel, 
then flip this spin (with probability one --- no rejection) 
and proceed to the next step.  
\item[$\circ$] One of the $4\ell$$-$$4$ spins on the 
boundary, then {\tt wait until\/} the {\it local\/} simulated 
virtual time 
of the next update is not greater than the virtual time 
on the neighboring PE(s).  
This spin {\it may or may not\/} flip: flip it using the 
flip probability of Eq.~(\ref{EIGlauber}) or 
Eq.~(\ref{EIMetrop}) and 
proceed to the next step.  
\end{itemize}
\item If the chosen spin was flipped, update the $10$ spin classes 
$n_i$.  
\item Add $\Delta t$ to the local virtual time.  
\item Use a random number ${\bar r}$ to determine the 
time of the new next update (in units of Monte Carlo steps --- mcs) 
by using the {\it local\/} time 
increment 
\begin{equation}
\label{EdtP} 
\Delta t = 
- {{\ell^2 \ln({\bar r})}\over{N_{\rm b}+\sum_{i=1}^{10}n_ip_i}}
\; .
\end{equation}
\end{itemize}
The performance of this algorithm with fixed $\ell$ 
as a function of $N_{\rm PE}$ 
is shown in 
Fig.~\ref{figT3Eup}.  For the parameters of the figure, 
the parallel $n$-fold way algorithm is almost twice as fast as 
the parallel standard dynamic Monte Carlo algorithm.  
For the parallel $n$-fold way algorithm it was found that 
the average utilization $\langle u\rangle$ was about ${1\over 2}$ 
(Fig.~\ref{figT3Eutil}).  
This utilization is lower than the utilization of the 
parallel standard Monte Carlo simulation 
(Fig.~\ref{figT3Eutil}).  Nevertheless, as seen in 
Fig.~\ref{figT3Eup} the $n$-fold way performs better in 
the number of spin flip trials per second per PE than does 
the parallel standard Monte Carlo algorithm.  
The introduction of the shield spins as an absorbing class 
limits the performance of the parallel $n$-fold way algorithm 
at low temperatures \cite{KORN99A}.  This is because at 
low temperatures the flip probabilities for each class, $p_i$, 
are small and the absorption almost always will take place 
in the class with the $N_{\rm b}$ shield spins.  In this case 
the average time step is limited by the block size, and 
is approximately 
$\langle\Delta t\rangle$$\approx$$\ell/4$ \cite{KORN99A}.  


\section*{6.0~Summary and Discussion}
\label{Sec-D}

Several advanced dynamic Monte Carlo algorithms have been 
introduced.  
All these algorithms preserve the underlying dynamic of the 
model --- they just implement it in a more 
intelligent fashion on the computer.  
One algorithm is the 
Monte Carlo with Absorbing Markov Chains (MCAMC) algorithm.  
The lowest-order, $s$$=$$1$, MCAMC algorithm corresponds to a 
discrete-time version of the $n$-fold way algorithm \cite{BORT75,NOVO95}.  
The other algorithm is the projective dynamics algorithm.  
We have shown that these advanced algorithms can be applied to 
models to study escape from metastable states.  
In all cases, the advanced algorithms do {\it not\/} alter the 
underlying dynamic of the model, but only implement the underlying 
dynamic in a different fashion.  
These algorithms can be {\it many\/} orders of magnitude faster 
than the standard Monte Carlo simulations.  Consequently, they 
allow simulations to be performed in parameter regimes it is 
impossible to study with standard algorithms.  

The MCAMC method was applied to a simple lattice model for 
motion of a random walker through a one-dimensional 
energy landscape in Sec.~4.0.  
This is a preliminary problem to 
using such methods to simulate motion of a particle through a 
free-energy landscape caused by a pore, such as 
ionic motion through a biological ion channel.  
It was shown that the $n$-fold way can be much faster 
than standard Monte Carlo.  The $n$-fold way algorithm, sometimes 
called event-driven simulation or a rejection-free algorithm, 
corresponds to $s$$=$$1$ MCAMC.  
In this model the higher $s$ MCAMC algorithms were introduced and 
found to sometimes be orders of magnitude faster 
than the $n$-fold simulations.  
For this simple model, the projective dynamics algorithm was found to 
be solvable for a finite lattice.  For more realistic simulations of 
motion through pores this will no longer be the case, but the 
projective dynamics algorithm is expected to again be 
applicable.  

Metastable decay of the Ising ferromagnet was illustrated using the 
MCAMC and projective dynamics algorithms.  
Again, the higher $s$ MCAMC algorithms were shown to often be many, many 
orders of magnitude faster than lower $s$ MCAMC algorithms or 
standard Monte Carlo.  
The disadvantage of the MCAMC algorithms is that they require a 
substantial amount of bookkeeping.  On the other hand, the 
projective dynamics algorithm can also be many orders of magnitude 
faster than conventional simulations, and it does not require very much 
bookkeeping.  In addition, the projective dynamics method can 
provide the magnetization of the metastable and stable states 
as well as the intervening saddle point.  
However, if higher moments of the distribution of 
escape times are desired, the projective dynamics algorithm 
only provides approximations.  
The MCAMC algorithms were found to work best at low temperatures 
in reasonably strong fields: near the single-droplet regime 
of the `metastability phase diagram'.  The 
projective dynamics algorithm was shown to also work in the 
multi-droplet regime.  

Although both the MCAMC and projective dynamics algorithms were 
initially designed for dynamic Monte Carlo simulations of 
models with discrete states, they have recently started to be 
generalized to spin models with continuous degrees of freedom.  
Here although there is an underlying spin dynamic 
\cite{SpinDPL} 
which is used for example in micromagnetic calculations 
\cite{MicroMag,BROW00}, 
in some cases it is still possible to ascribe a physical 
meaning to a Monte Carlo move \cite{NOWA00}.  
For the classical Heisenberg model recent exploratory 
efforts to apply the $n$-fold way 
\cite{MUNO00A,MUNOPRL} and the projective dynamics methods 
\cite{MITC00} have been encouraging.  

The advanced dynamic Monte Carlo algorithms presented here 
have the goal of accelerating the simulation of dynamics 
for physical systems.  Methods with the same goal for 
molecular dynamics simulations have also recently been 
advanced by Voter.  
These include the 
hyperdynamics \cite{VOTE97}, 
parallel replica dynamics \cite{VOTE98}, 
and temperature-extrapolated dynamics \cite{VOTE00}.  
These three methods can be used either individually or 
in combination for molecular dynamics simulations.  They all 
assume that transition-state theory \cite{JOHN66,BAER96} 
is satisfied.  This assumption is related to the 
assumption in the projected dynamics that the ratio of growing and 
shrinking probabilities is given by Boltzmann weights.  
As shown in \cite{LEE95} for the Ising model, such an assumption works 
reasonably well in the single-droplet regime but fails in the 
multidroplet regime.  Furthermore, in Ref.~\cite{LEE95} the prefactor 
of the nucleation rate was shown to depend on the explicit dynamic 
used, even though the dynamics all satisfied assumptions similar to 
transition-state theory.  It can be anticipated that only in the 
regimes where the growth phase is unimportant will the 
transition-state theory assumptions be satisfied, and only there will 
Voter's advanced molecular dynamics algorithms give correct results.  
Conversely, neither the MCAMC nor the projective dynamics methods 
presented here depend on any assumptions related to 
transition-state theory.  They are just different ways to implement 
the original dynamic.  Hence these advanced dynamic Monte Carlo algorithms 
give the correct result.  

Another recent dynamic Monte Carlo algorithm 
called Transition Matrix Monte Carlo (TMMC) has been proposed by 
Wang, Tay, and Swendsen \cite{WANG99B}.  
The idea is very similar to that of projective dynamics 
\cite{KOLE98A,KOLE98B}.  The underlying $2^N$$\times$$2^N$ 
Markov matrix of the $N$ spin Ising model is lumped 
using equations identical to Eq.~(\ref{EIPD2}) and (\ref{EIPD3}) 
but with energy rather than magnetization as the lumping variable.  
Wang et al.\ study the square-lattice Ising model at $T_{\rm c}$ in zero 
field.  
The lumped Markov matrix they obtain is not tridiagonal like 
the projective dynamics Markov matrix in Sec.~5.4, 
so our proof that the dynamic is unaffected by the lumping 
is no longer applicable.  
In fact, they obtain dynamic relaxation times 
different from those of the unlumped Markov matrix.  
Consequently, the TMMC method must alter the underlying 
dynamics of the model.  
It would be interesting to project onto both slow variables, 
$M$ and energy, for the Ising model \cite{SHTE97,SHTE99}.  
This can be done \cite{LEE95} 
using the formal relationship between the projective dynamics and 
the Nakjima-Zwanzig \cite{NAKA58,ZWAN60} projection-operator formalism 
for the master equation, which is equivalent to Mori's \cite{MORI65} 
projection-operator formalism for the equations of motion of 
observables \cite{GRAB77,GRAB82}.  Projecting onto both slow variables 
should make the underlying Markov matrix closer to weakly lumpable.  

In summary, the projective dynamics and 
MCAMC algorithms allow dynamic Monte Carlo simulations to be 
performed 
cheaper, better, faster --- 
without changing the underlying dynamic.  
They should be used in all situations where long time 
dynamic Monte Carlo simulations 
are required to understand the underlying physics of the 
discrete-state-space model 
under investigation.  

\centerline{\bf Acknowledgments}

The author gratefully acknowledges discussions with 
M.\ Kolesik, 
G.\ Korniss, 
J.~Lee, 
S.~J.\ Mitchell, 
J.~D.\ Mu{\~n}oz, 
R.~A.\ Ramos, 
H.~L.\ Richards, 
and 
P.~A.\ Rikvold.  
This work was supported by 
the U.S.\ National Science Foundation Grant No.~DMR-9871455.  
This research used the resources of 
the National Energy Research Scientific Computer Center 
(NERSC), which is supported by the Office of Energy Research of the 
US DOE under Contract No.~DE-AC03-76SF00098.  

\clearpage

\begin{thebibliography}{100}
\bibitem{BKTERR}
There is a misprint near the bottom of page~13 in the
original $n$-fold way paper of 
Ref.\ \protect\cite{BORT75}. 
The randomly chosen integer should be in the range 
$[1,n_i]$ to find LCH, not $[1,m_i]$. 
I thank B.~M.\ Gorman for pointing out this misprint.
\end{thebibliography}

\clearpage

\clearpage

\clearpage

\clearpage

\end{document}